\documentclass[journal=jctcce,manuscript=article]{achemso}
\usepackage{booktabs}
\usepackage{multirow}
\usepackage{graphicx}
\usepackage{dcolumn}
\usepackage{bm}
\usepackage{enumitem}
\usepackage{hyperref}
\usepackage{bigints}
\usepackage{rotating}
\usepackage{siunitx}
\usepackage[autostyle]{csquotes}
\usepackage[table,xcdraw]{xcolor}
\definecolor{codegreen}{rgb}{0,0.6,0}
\definecolor{codegray}{rgb}{0.5,0.5,0.5}
\definecolor{codepurple}{rgb}{0.58,0,0.82}
\definecolor{backcolour}{rgb}{0.95,0.95,0.92}
\usepackage[T1]{fontenc}
\usepackage[english]{babel}
\MakeOuterQuote{"}
\usepackage{siunitx}
\usepackage[version=4]{mhchem} 
\usepackage[acronym]{glossaries}

  \newcommand*{\citen}{}
\DeclareRobustCommand*{\citen}[1]{%
  \begingroup
    \romannumeral-`\x 
    \setcitestyle{numbers}%
    \cite{#1}%
  \endgroup
}

\makeatletter
\newcommand\reaction@[1]{\begin{equation}\ce{#1}\end{equation}}
\newcommand\reaction@nonumber[1]%
{\begin{equation*}\ce{#1}\end{equation*}}
\newcommand\reaction{\@ifstar{\reaction@nonumber}{\reaction@}}
\makeatother

\usepackage{braket}

\newcommand{\mr}{\mathrm}

\newcommand{\bk}[2]{\langle #1 \vert #2 \rangle}
\newcommand{\ev}[1]{\langle #1 \rangle}
\newcommand{\expv}[3]{\langle #1 \vert #2 \vert #3 \rangle}

\newacronym{sic}{SIC}{self-interaction correction}
\newacronym{lrdft}{LR-TDDFT}{linear-response time-dependent density functional theory}
\newacronym{paw}{PAW}{projector augmented wave}
\newacronym{tdm}{TDM}{transition dipole moment}
\newacronym{dm}{DM}{dipole moment}
\newacronym{1rtdm}{1-RTDM}{one-particle reduced transition density matrix}
\newacronym{1rdm}{1-RDM}{one-particle reduced density matrix}
\newacronym{ao}{AO}{atomic orbital}
\newacronym{mo}{MO}{molecular orbital}
\newacronym{lcao}{LCAO}{linear combination of atomic orbitals}
\newacronym{rsg}{RSG}{real-space grid}
\newacronym{pw}{PW}{plane-wave}
\newacronym{xc}{xc}{exchange-correlation}
\newacronym{ks}{KS}{Kohn-Sham}
\newacronym{doo}{DOO}{direct orbital optimization}
\newacronym{oo}{OO}{orbital-optimized}
\newacronym{mom}{MOM}{maximum overlap method}
\newacronym{si}{SI}{Supporting Information}
\newacronym{cbs}{cbs}{complete basis set}
\newacronym{fc}{FC}{Franck-Condon}
\newacronym{tbe}{TBE}{theoretical best estimate}
\newacronym{homo}{HOMO}{highest occupied molecular orbital}
\newacronym{lumo}{LUMO}{lowest unoccupied molecular orbital}
\newacronym{dft}{DFT}{density functional theory}
\newacronym{aug}{aug}{\mbox{aug-cc-pVDZ+sz}}
\newacronym{daug}{d-aug}{\mbox{d-aug-cc-pVDZ+sz}}
\newacronym{tda}{TDA}{Tamm-Dancoff approximation}
\newacronym{fwhm}{FWHM}{full width at half maximum}
\newacronym{gpaw}{GPAW}{grid-based project augmented wave}

\title{Excited-state Properties Beyond the Excitation Energy from
Orbital-Optimized Density Functional Calculations II: Absorption Spectra}

\author{Lorenzo Restaino}
\email{lorenzo@hi.is}
\affiliation{Science Institute and Faculty of Physical Sciences, University of Iceland, 107 Reykjavík, Iceland}
\author{Diego Llorena Prieto}
\affiliation{Science Institute and Faculty of Physical Sciences, University of Iceland, 107 Reykjavík, Iceland}
\author{Jukka John}
\affiliation{Science Institute and Faculty of Physical Sciences, University of Iceland, 107 Reykjavík, Iceland}
\author{Yorick Leonard Adrian Schmerwitz}
\affiliation{Max-Planck-Institut f\"ur Kohlenforschung, 45470 M\"ulheim an der Ruhr, Germany}
\author{Elvar Örn Jónsson}
\affiliation{Science Institute and Faculty of Physical Sciences, University of Iceland, 107 Reykjavík, Iceland}
\author{Gianluca Levi}%
\email{gianluca.levi@units.it}
\affiliation{Department of Chemical and Pharmaceutical Sciences, University of Trieste, 34127 Trieste, Italy}
\affiliation{Science Institute and Faculty of Physical Sciences, University of Iceland, 107 Reykjavík, Iceland}
\begin{document}

\begin{abstract}
Previous benchmarks of orbital-optimized density functional calculations of transition intensities in molecules have primarily been restricted to the lowest electronic excitation. Here, the performance of this variational approach is assessed across several valence and Rydberg excited states of small molecules, calculating the absorption spectra up to 10~eV. A plane-wave basis set is employed to accurately describe diffuse Rydberg excitations, while the transition dipole moment is evaluated from nonorthogonal Kohn--Sham determinants within the projector augmented wave formalism. Comparison with higher-level coupled-cluster results shows that orbital-optimized calculations provide a good description of oscillator strengths for states with predominantly single-configurational character, even with the generalized gradient approximation functional PBE. The inclusion of exact exchange and self-interaction correction further improves the results, with the latter yielding the smallest errors (mean absolute relative error of $\sim$25\%). In contrast, large errors for all functionals are found for multi-configurational states, due to an inability of the approach to describe the multi-configurational character. Nonorthogonality between independently optimized states has previously been raised as a possible source of error in transition properties. In the present benchmark, no correlation of the nonorthogonality between the ground and excited states with the error on the transition intensity is found. These results establish the performance of orbital-optimized density functional calculations for absorption spectra of small molecules and highlight the need for extensions that combine state-specific orbital relaxation with a multi-configurational treatment.
\end{abstract}

\maketitle
\section{Introduction}
Electronic excited states of molecules, including low-lying ones, can have a Rydberg character, 
in which an electron has a spatially diffuse density extending far from the molecule. In absorption spectra, molecular Rydberg states often appear as series of transitions converging toward the ionization threshold, with spacings and intensities governed by the interaction between the diffuse excited electron and the molecular core~\cite{Reisler2009-hz,Softley2004-gl,Kuthirummal2003-pa,Sandorfy1999-bv,Merkt1997-na}. Such states are generally less bright than valence states. Nevertheless, the intensity of transition to Rydberg states can increase when the excited electron has only moderately diffuse density or when the Rydberg state interacts and mixes with nearby valence states of the same symmetry~\cite{Zheng2016-vh,Escure2009-pz,Reisler2009-hz}. In this case, the resulting adiabatic states acquire mixed Rydberg–valence character, and the oscillator strengths of transitions to these states can be substantially larger than those of pure Rydberg states. Buenker and Peyerimhoff~\cite{Buenker1975-qt} describe this in terms of mixed valence–Rydberg wave functions, where the oscillator strength is dominated by the valence contribution, while the Rydberg contribution is reflected more strongly in the spatial extent of the excited state.

The calculation of absorption spectra involving Rydberg transitions remains a challenge for many electronic structure methods. Correlated wave function approaches can provide an accurate description, but they are computationally demanding and require a careful choice of the active space, especially when several diffuse Rydberg orbitals and nearby valence and charge-transfer states must be described simultaneously~\cite{Kaufold2023-mz,Zobel2017-cq,Veryazov2011-kj,Serrano-Andres1996-rm,Serrano-Andres1996-vk}. \Gls*{lrdft}~\cite{Casida1995, Runge1984} is often used as an affordable method for calculating absorption spectra, but Rydberg states remain difficult to describe for standard implementations in combination with the adiabatic approximation. The limitation here is that the effective potential of approximate \gls*{ks}~\cite{Kohn1965, Hohenberg1964} \gls*{xc} functionals is upshifted and fails to provide the correct asymptotic $-1/r$ decay of the tail, where $r$ denotes the distance from an atom, which affects the description of diffuse Rydberg states within \gls*{lrdft}~\cite{VanMeer2014}. As a result, Rydberg states may be placed at incorrect energies and improperly mixed with nearby valence and charge-transfer states~\cite{Selenius2024-rk,Li2015-yj,Seidu2015-fz,VanMeer2014,Cheng2008,Peach2008,Tozer2000-kw,Casida1998}, which in turn affects the quality of the computed spectra. The choice of the basis set is also critical. When using common \gls*{lcao} basis sets, if the basis functions are not sufficiently diffuse, the spatial extent of Rydberg states is artificially confined. At the same time, inclusion of a large number of diffuse functions can lead to linear dependence in the calculations~\cite{Palmer2018-jv,Kaufmann1989-lb}.

\Gls*{oo} density functional calculations~\cite{Restaino2026-yg, Herbert2023, Hait2021} provide an alternative, time-independent route in which each excited state is found variationally  rather than described within the linear-response from the ground state. In this framework, excited states are obtained through state specific orbital optimization, and therefore orbital relaxation is included explicitly. In fully variational \gls*{oo} formulations~\cite{Restaino2026-yg}, where no constrains on the obtained solutions are imposed, excited states are stationary points with nonaufbau occupation on the electronic energy surface given by an energy functional of the density. Since these solutions are typically saddle points and not minima, as the ground state, tailored wave function optimization techniques~\cite{Qin2026, Schmerwitz2026, Bogo2025, Schmerwitz2023, Levi2020-nz, Carter2020, Levi2020farad, Hait2020, Barca2018, Gilbert2008, Cheng2008} are required.

Several studies have shown that \gls*{oo} density functional approaches can improve the excitation energy of Rydberg states compared to conventional \gls*{lrdft}~\cite{Sigurdarson2023-do,Seidu2015-fz,Yang2011-rm} calculations. However, far less is known about the performance of \gls*{oo} calculations with respect to the intensity of transitions and optical spectra of molecules, especially when they involve Rydberg excitations. Approximating the true wave function with the \gls*{ks} wave function to evaluate the \gls*{tdm} has been shown to lead to a reasonable approximation of transition properties~\cite{Yang2026-et, Qin2026-mf, Toffoli2022, Bourne-Worster2021-hb,Hait2020-il,Hait2020-fy, Gilbert2008}. However, this approach comes with an additional complication. Since each electronic state is described by its own set of optimized molecular orbitals, the states are generally not orthogonal to each other, and matrix elements must be evaluated between sets of mutually nonorthogonal orbitals.

Gilbert \textit{et al.} were the first to point out possible issues in evaluating transition properties for nonorthogonal \gls*{oo} excited states in their original \gls*{mom} paper~\cite{Gilbert2008-ye}. They examined whether nonzero overlaps between the OO states could artificially inflate the calculated \glspl*{tdm}. Their analysis shows that even when the ground and excited states share the same irreducible representation and have finite overlap, the resulting oscillator strengths still fall within the range obtained from configuration interaction singles (CIS) and \gls*{lrdft}. They also report that \gls*{oo} \gls*{ks} calculations with the B3LYP functional yield \glspl*{tdm} very similar to OO Hartree-Fock calculations, even though the overlaps are considerably smaller for the former. Bourne Worster et al.~\cite{Bourne-Worster2021-hb} pointed out that when the \gls*{tdm} is evaluated using only the electronic part of the dipole moment operator, the \gls*{tdm} becomes origin dependent. Including the nuclear contribution, which is an integral part of the dipole moment operator, preserves translational invariance for charge neutral systems, although the origin dependence still affects charged systems. To circumvent the issue of nonorthogonality and lack of translational invariance, Bourne Worster et al. used a symmetric orthogonalization procedure to orthogonalize the \gls*{oo} states. This strategy was tested on a large set of molecules, but limited to only one excited state per molecule, specifically the HOMO-LUMO transition. The authors found that oscillator strengths of neutral molecules obtained using symmetric orthogonalization and only the electronic part of the dipole moment operator closely match those obtained from nonorthogonal states when including the nuclear dipole moment. Sinyavskiy \textit{et al.}~\cite{Sinyavskiy2025} proposed an alternative strategy that avoids evaluating the \gls*{tdm} directly between nonorthogonal \gls*{oo} states. The optimized excited state is recast in terms of singly excited configurations built from the ground-state orbitals by projection, and the \gls*{tdm} is evaluated within this common set of orbitals. Although computationally convenient, this procedure replaces the original \gls*{oo} excited state with a representation in terms of ground-state orbitals, thereby losing part of the state specific orbital relaxation effects encoded in the optimized excited state. A recent work by Shen \textit{et al.}~\cite{Yang2026-et} extended the calculation of oscillator strengths for \gls*{oo} nonorthogonal states to the velocity gauge and applied the methodology to the same set of molecular excited states considered by Bourne Worster \textit{et al.}~\cite{Bourne-Worster2021-hb}. Since the transition intensity is formulated in terms of the momentum operator rather than the dipole operator, this approach does not suffer from origin dependence, even for charged systems. The authors report that the resulting oscillator strengths are close to those obtained in the length gauge with symmetric orthogonalization. However, as in other works~\cite{Vandaele2022-zl,Toffoli2022-ud,Bourne-Worster2021-hb}, the study is limited to a single HOMO-LUMO excitation for each molecule.

Despite these recent developments, it remains unclear to what extent the finite overlap between independently optimized states in OO calculations affects the accuracy of the calculated \glspl*{tdm} and the resulting optical intensities. Moreover, not much is currently known about how \gls*{oo} density functional methods perform for oscillator strengths of excitations beyond the HOMO-LUMO transition. This limitation partly reflects the fact that conventional self-consistent field (SCF) procedures are not designed for convergence to saddle points and, as a result, frequently struggle in OO calculations of excited states.

In this work, \gls*{oo} density functional calculations are used to predict optical intensities of valence as well as Rydberg excited states above the lowest-energy excitation for several neutral small molecules using a robust direct orbital optimization approach~\cite{Ivanov2021, Levi2020-nz} for obtaining the excited-state solutions. The \gls*{oo} calculations are carried out for excited states of water (H$_2$O), formaldehyde (CH$_2$O), ammonia (NH$_3$), methanol (CH$_3$OH), and ethylene (C$_2$H$_4$), extending up to 10~eV above the ground state. L{\"o}wdin's rules for nonorthogonal determinants~\cite{Figari1985-ds,Lowdin1955-ve} are employed to calculate the transition matrix elements directly using the nonorthogonal states. Although these rules are well established, to the best of our knowledge, this work presents for the first time their formulation and application within the \gls*{paw} formalism~\cite{Blochl1994-gk} and implements the resulting expressions for both LCAO and \gls*{pw} representations of the orbitals in the \gls*{gpaw} software~\cite{Mortensen2024-ji,Mortensen2005-nw}. The oscillator strengths are evaluated using several \gls*{xc} functionals, namely the generalized gradient approximation (GGA) functional PBE~\cite{Perdew1997-rt,Perdew1996-il}, the hybrid functional PBE0~\cite{Adamo1999}, and PBE with the explicit Perdew–Zunger \gls*{sic}~\cite{Perdew1981-oz} employing a globally scaled \gls*{sic}. The inclusion of \gls*{sic} restores the correct asymptotic $-1/r$ behavior of the effective potential~\cite{Sigurdarson2023-do,Levi2020-nz,Melander2016-qp,Chai2008-rw}, which is particularly relevant for Rydberg states. Calculations using \gls*{lrdft} with the PBE and PBE0 functionals are also performed. The OO and \gls*{lrdft} results are then compared to higher-level correlated wave function calculations performed with sufficiently diffuse \gls*{lcao} basis sets. 

An accompanying article~\cite{Restaino2026-iv} assesses the performance of \gls*{oo} density functional calculations for excited-state dipole moments for a set of molecules with valence and Rydberg states, including some of those investigated here.

\section{Methodology}
\subsection{Orbital-optimized excited state calculations}
The excited states found in fully variational \gls*{oo} calculations correspond to stationary points on the energy surface given by the variation of the energy as a function of the electronic degrees of freedom. In OO KS density functional calculations, these solutions also satisfy the KS equations. Unlike the ground-state solution, the OO excited states are usually saddle points rather than minima on the electronic energy surface~\cite{Restaino2026-yg, Schmerwitz2023}. As a consequence, conventional SCF algorithms based on repeated diagonalization of the Hamiltonian matrix may collapse to a lower-energy solution or fail to converge altogether~\cite{Qin2026-mf, Hait2021, Levi2020-nz, Carter2020, Hait2020}.

This challenge is addressed here using a direct optimization strategy~\cite{Schmerwitz2026, Schmerwitz2023, Ivanov2021, Levi2020-nz, Levi2020farad}. Given an initial set of orthonormal orbitals
$\boldsymbol{\psi}_0=\{\psi_i^0(\mathbf r)\,|\,1\le i\le N_{\mr{orb}}\}$,
the optimal orbitals that make the energy stationary are obtained through repeated application of a unitary transformation parametrized as the exponential of an anti-Hermitian matrix~\cite{Head-Gordon1988},
\begin{equation}
    \boldsymbol{\psi}=\boldsymbol{\psi}_0e^{\boldsymbol{\kappa}}, \qquad
    \boldsymbol{\kappa}=-\boldsymbol{\kappa}^{\dagger}.
\end{equation}
In general, an excited state can then be located by solving a nested variational problem in which the energy is made stationary with respect to $\boldsymbol{\kappa}$ and minimized with respect to $\boldsymbol{\psi}_0$~\cite{Ivanov2021},
\begin{equation}
    \underset{\boldsymbol{\psi}}{\mathrm{stat}}\,
    E[\boldsymbol{\psi}]
    =
    \underset{\boldsymbol{\psi}_0}{\mathrm{min}}\,
    \underset{\boldsymbol{\kappa}}{\mathrm{stat}}\,
    E[\boldsymbol{\psi}_0 e^{\boldsymbol{\kappa}}].
\end{equation}
The calculations reported here use both \gls*{pw} and \gls*{lcao} representations of the orbitals, employing direct optimization as implemented in the GPAW program~\cite{Ivanov2021, Levi2020-nz}.

\subsection{Self-interaction correction}\label{sec:sic}
In approximate KS density functional calculations, the electron--electron repulsion is partly affected by a self-interaction error. This originates from the Hartree energy, which is evaluated from the total electron density and therefore contains the Coulomb interaction of an electron density with itself. For the exact xc functional, this spurious term is removed exactly. However, local and semi-local approximations cannot fully cancel it, since the \gls*{sic} has a nonlocal character.

In the present work, \gls*{sic} calculations using the Perdew--Zunger formulation~\cite{Perdew1981-oz} are performed. In this approach, the self-Hartree and self-exchange--correlation contributions are subtracted one by one for each occupied orbital density, giving the total \gls*{sic} energy
\begin{equation}\label{eq:sic}
E_{\mathrm{SIC}}[
n_1,n_2,\ldots,n_N]
=
E_{\mathrm{KS}}[n]
-
\alpha \sum_i^N
\left(
E_{\mathrm H}[n_i]
+
E_{\mathrm{XC}}[n_i, 0]
\right),
\end{equation}
where $E_{\mathrm{KS}}$ is the approximate KS energy, $E_{\mathrm H}$ is the Hartree energy, $E_{\mathrm{XC}}$ is the xc energy, $n$ is the total density, $n_i$ is the density of orbital $i$, and $\alpha$ is a global scaling factor.  The latter is used because the unscaled Perdew--Zunger correction often overcorrects approximate density functionals. Scaled corrections, commonly with $\alpha=1/2$, have been shown to improve results for the atomization and excitation energy of molecules~\cite{Ivanov2021, Lehtola2016b}, as well as band gaps of solids ~\cite{Gudmundsdottir2015}. Because the correction is applied to the individual orbital densities, the resulting functional depends explicitly on the set of occupied orbitals and not only on the total density. As a result, the \gls*{sic} energy is not invariant under unitary transformation within the space of occupied orbitals, and the direct optimization approach described in the previous section must take into account rotations among occupied orbitals~\cite{Ivanov2021, Schmerwitz2024}. For excited-state calculations, this provides a fully variational optimization of the \gls*{sic} orbitals, accounting simultaneously for the orbital relaxation of the excitation and the orbital localization driven by the \gls*{sic}.

\subsection{Intensity of electronic transitions}
The intensity of the electronic transitions used to obtain the absorption spectra is evaluated from the corresponding oscillator strengths within the electric dipole approximation. In the length gauge, the dimensionless oscillator strength associated with a transition between states $\Psi^k$ and $\Psi^{k^\prime}$ is given by
\begin{equation}\label{eq:osc_strength}
f^{kk^\prime}
=
\frac{2}{3}\Delta E^{kk^\prime}
\left|\bm{\mu}^{kk^\prime}\right|^2 ,
\end{equation}
where $\Delta E^{kk^\prime}=E^{k^\prime}-E^k$ is the energy separation between the states and $\bm{\mu}^{kk^\prime}$ is the \gls*{tdm} evaluated using the dipole moment operator. For a molecular system of $N$ electrons and $M$ nuclei, the total dipole moment operator in atomic units is
\begin{align}
\hat{\bm{\mu}}
=
-\sum_{p}^{N}\mathbf{r}_p
+\sum_{a}^{M}\mathcal{Z}_a\mathbf{R}_a ,
\end{align}
where $\mathbf{r}_p$ denotes the position of electron $p$, while $\mathcal{Z}_a$ and $\mathbf{R}_a$ are the charge and position of nucleus $a$, respectively. The \gls*{tdm} between states $k$ and $k^\prime$ can therefore be written as
\begin{equation}
\bm{\mu}^{kk^\prime}
=
-\left\langle\Psi^k \left\vert \sum_{p}^{N}\mathbf{r}_p \right\vert \Psi^{k^\prime}\right\rangle
+
\bk{\Psi^k}{\Psi^{k^\prime}}
\sum_{a}^{M}\mathcal{Z}_a\mathbf{R}_a .
\label{eq:trans_dipole}
\end{equation}
For orthonormal states, $\bk{\Psi^k}{\Psi^{k^\prime}}=\delta_{kk^\prime}$, and the nuclear contribution vanishes. In the present work, however, the independently optimized states are generally nonorthogonal, and the nuclear term does not cancel. Inclusion of this term ensures that the \gls*{tdm} is translationally invariant for a neutral molecule. In contrast, unless the states are orthogonalized, neglecting the nuclear term leads to an origin-dependent \gls*{tdm} and, consequently, an origin-dependent oscillator strength.

\subsubsection{Calculation of matrix elements between nonorthogonal states}
Since OO states are made of sets of mutually nonorthogonal orbitals, the matrix elements in eq\ \eqref{eq:trans_dipole} cannot be evaluated using the Slater--Condon rules. Here, we use L{\"o}wdin's rules for nonorthogonal determinants~\cite{Figari1985-ds,Lowdin1955-ve}, which have already been applied to the \gls*{oo} calculation of the \gls*{tdm} in previous studies~\cite{Toffoli2022-ud,Bourne-Worster2021-hb}. Related treatments of independently optimized, nonorthogonal wave functions have also been used to compute photoelectron intensities, including double-core-hole spectra obtained with selected configuration interaction~\cite{Ferte2020}.

The matrix elements of a one-body operator $\hat O$ can be written as~\cite{Figari1985-ds,Lowdin1955-ve}
\begin{equation}\label{Lowdin_NOCI}
    \langle \Psi ^k | \hat O | \Psi ^{k^{\prime}}\rangle
    =
    \sum_{ij}
    O_{ij}^{kk^{\prime}}
    \,
    \mathrm{cof}(\mathbf{S}^{kk^{\prime}})_{ij}
\end{equation}
where $\ket{\Psi ^k}$ is a Slater determinant built from occupied spin orbitals $\{\psi_i^k\,|\,1\le i\le N_{\mr{orb}}\}$ and
\begin{align}
    O_{ij}^{kk^{\prime}}
    &=
    \expv{\psi_i^k}{\hat O}{\psi_j^{k^{\prime}}},
    \\
    S_{ij}^{kk^{\prime}}
    &=
    \bk{\psi_i^k}{\psi_j^{k^\prime}}.
\end{align}
The cofactor of the overlap matrix can be expressed in terms of its adjugate as
\begin{equation}
\operatorname{cof}\!\left(\mathbf{S}^{kk^\prime}\right)_{ij}
=
\operatorname{adj}\!\left(\mathbf{S}^{kk^\prime}\right)_{ji}
=
\det\!\left(\mathbf{S}^{kk^\prime}\right)
\left(\mathbf{S}^{kk^\prime}\right)^{-1}_{ji},
\end{equation}
where the determinant of the orbital overlap matrix is equal to the overlap between the two Slater determinants,
$\det\!\left(\mathbf{S}^{kk^\prime}\right)=\bk{\Phi^k}{\Phi^{k^\prime}}$.
In some previous works, eq\ \eqref{Lowdin_NOCI} is written using the adjugate of the overlap matrix. However, as already pointed out in ref.~\citen{Lemke2024-tg}, the transpose of the adjugate, i.e., the cofactor matrix, should be used instead. For spin-unrestricted calculations, as adopted in the present work, the overlap matrix is block diagonal,
\begin{equation}
\mathbf{S}^{kk'}
=
\begin{pmatrix}
\mathbf{S}^{kk'}_{\alpha} & 0\\
0 & \mathbf{S}^{kk'}_{\beta}
\end{pmatrix},
\end{equation}
and the cofactor matrix of $\mathbf{S}^{kk'}$ becomes
\begin{equation}\label{cofactor_unrestr}
\mathrm{cof}(\mathbf{S}^{kk'})
=
\begin{pmatrix}
\det(\mathbf{S}^{kk'}_{\beta})\,\mathrm{cof}(\mathbf{S}^{kk'}_{\alpha}) & 0\\
0 & \det(\mathbf{S}^{kk'}_{\alpha})\,\mathrm{cof}(\mathbf{S}^{kk'}_{\beta})
\end{pmatrix}.
\end{equation}
Using eq\ \eqref{cofactor_unrestr}, the matrix elements between unrestricted nonorthogonal determinants are written as
\begin{equation}\label{eq:Lowdin_NOCI_unrestr}
 \langle \Psi^k | \hat O | \Psi^{k^{\prime}}\rangle
 =
 \det(\mathbf{S}^{kk^\prime}_{\beta})
 \sum_{ij\in\alpha}
 O_{ij}^{kk^{\prime}}
 \,
 \mathrm{cof}(\mathbf{S}^{kk^{\prime}}_{\alpha})_{ij}
 +
 \det(\mathbf{S}^{kk^\prime}_{\alpha})
 \sum_{ij\in\beta}
 O_{ij}^{kk^{\prime}}
 \,
 \mathrm{cof}(\mathbf{S}^{kk^{\prime}}_{\beta})_{ij}.
\end{equation}

Therefore, using the matrix elements of the dipole moment operator between nonorthogonal OO states, the expression of the OO \gls*{tdm} in eq\ \eqref{eq:trans_dipole} becomes
\begin{align}\label{eq:Lowdin_NOCI_unrestr_final}
 \langle \Psi^k | \hat{\bm{\mu}} | \Psi^{k^{\prime}}\rangle
 =&
 -\det(\mathbf{S}^{kk^\prime}_{\beta})
 \sum_{ij\in\alpha}
 \mathbf{r}_{ij}^{kk^{\prime}}
 \,
 \mathrm{cof}(\mathbf{S}^{kk^{\prime}}_{\alpha})_{ij}
 -\det(\mathbf{S}^{kk^\prime}_{\alpha})
 \sum_{ij\in\beta}
 \mathbf{r}_{ij}^{kk^{\prime}}
 \,
 \mathrm{cof}(\mathbf{S}^{kk^{\prime}}_{\beta})_{ij}
 \nonumber\\
 &
 +\det(\mathbf{S}^{kk^\prime}_{\alpha})
 \det(\mathbf{S}^{kk^\prime}_{\beta})
 \sum_{a}^{M}
 \mathcal{Z}_{a}
 \mathbf{R}_a,
\end{align}
where $\mathbf{r}_{ij}^{kk^{\prime}} = \expv{\psi_i^k}{\mathbf{r}}{\psi_j^{k^{\prime}}}$.

An equivalent formulation can be obtained by transforming the occupied
orbitals of the two states to a corresponding-orbital or biorthogonal
representation~\cite{Malmqvist1986-jl}, using, e.g., a singular value decomposition
that makes the orbital overlap matrix diagonal. The formulation remains well
defined even when the overlap matrix is singular, while the L{\"o}wdin's formula using the adjugate 
cannot be used when the overlap matrix is singular. In the biorthogonal representation,
matrix elements can also be evaluated using generalized Slater--Condon
rules~\cite{Burton2021}. This can provide a useful alternative for OO density functional calculations.

\subsubsection{\Acrlong*{tdm} in the projector augmented wave formalism}
Here, we present the evaluation of the transition dipole moment between nonorthogonal \gls*{oo} states within the \gls*{paw} formalism~\cite{Blochl1994-gk}.

Within the \gls*{paw} approach, the orbitals of electronic state $k$, $\ket{\psi_i^k}$, are related to the smooth pseudo orbitals optimized in the variational problem, $\ket{\tilde{\psi}_i^k}$, through the linear transformation operator $\hat{\mathcal T}$,
\begin{align}
\label{eq:paw_transform}
\ket{\psi_i^k} &= \hat{\mathcal T} \ket{\tilde{\psi}_i^k}\nonumber \\
\hat{\mathcal T} &= 1 + \sum_a\sum_n \left( \ket{\phi_n^a} - \ket{\tilde{\phi}_n^a} \right) \bra{\tilde p_n^a},
\end{align}
where $\ket{\tilde p_n^a}$, $\ket{\phi_n^a}$, and $\ket{\tilde{\phi}_n^a}$ are atom-centered projector functions, partial waves, and pseudo partial waves, respectively, and $a$ runs over the atoms. The transformation reconstructs the oscillatory behavior of the wave function in the vicinity of the nuclei within atom-centered augmentation spheres, while outside these regions, the orbitals and the pseudo orbitals are identical.

Within the PAW formalism with the frozen-core approximation for the orbitals, L{\"o}wdin's expression of the transition dipole moment between two spin-unrestricted nonorthogonal determinants is (see Appendix~\ref{sec:paw_nonorthogonal_matrix_elements} for a derivation)
\begin{align}
\label{eq:Lowdin_NOCI_unrestr_final_paw}
\expv{\Psi^k}{\hat{\bm{\mu}}}{\Psi^{k^\prime}}
={}&
-\det\!\left(\mathbf S_{\beta}^{kk^\prime}\right)
\sum_{ij\in\alpha}
\mathbf r_{ij}^{kk^\prime}
\,
\mathrm{cof}\!\left(\mathbf S_{\alpha}^{kk^\prime}\right)_{ij}
\nonumber\\
&-
\det\!\left(\mathbf S_{\alpha}^{kk^\prime}\right)
\sum_{ij\in\beta}
\mathbf r_{ij}^{kk^\prime}
\,
\mathrm{cof}\!\left(\mathbf S_{\beta}^{kk^\prime}\right)_{ij}
\nonumber\\
&+
\det\!\left(\mathbf S_{\alpha}^{kk^\prime}\right)
\det\!\left(\mathbf S_{\beta}^{kk^\prime}\right)
\sum_a
\left(
\mathcal Z_a-N_{\mathrm{core}}^a
\right)
\mathbf R_a,
\end{align}
where $N_{\mathrm{core}}^a$ is the number of frozen-core electrons for atom $a$, and it has been assumed that the frozen-core densities are spherical and centered on the atoms.  When the core orbitals and the partial waves form an orthonormal basis, as it is typically the case since they are obtained from the same atomic KS calculation~\cite{Mortensen2024-ji, Blochl2002}, the one-electron matrix elements of the position operator appearing in eq\ \eqref{eq:Lowdin_NOCI_unrestr_final_paw} are only the PAW-corrected matrix elements between valence orbitals,
\begin{align}
\mathbf r_{ij}^{kk^\prime}
=
\expv{\psi_i^k}{\mathbf r}{\psi_j^{k^\prime}}
={}&
\expv{\tilde{\psi}_i^k}{\mathbf r}{\tilde{\psi}_j^{k^\prime}}
+
\sum_a\sum_{nm}
\left[
\expv{\phi_n^a}{\mathbf r}{\phi_m^a}
-
\expv{\tilde{\phi}_n^a}{\mathbf r}{\tilde{\phi}_m^a}
\right]
D_{nm,ij}^{a,kk^\prime},
\label{eq:paw_position_matrix_element}
\end{align}
where
\begin{equation}
D_{nm,ij}^{a,kk^\prime}
=
P_{in}^{a,k*}P_{jm}^{a,k^\prime},
\qquad
P_{in}^{a,k}
=
\bk{\tilde p_n^a}{\tilde{\psi}_i^k},
\end{equation}
and $D_{nm,ij}^{a,kk^\prime}$ is an atomic transition density matrix. The second term in eq\ \eqref{eq:paw_position_matrix_element} corresponds to the PAW correction. Finally, the last term in the expression of the TDM in the PAW formalism, eq\ \eqref{eq:Lowdin_NOCI_unrestr_final_paw}, includes a contribution due to the frozen core electrons, which replaces the nuclear charge $\mathcal Z_a$ by the effective valence charge $\mathcal Z_a-N_{\mathrm{core}}^a$.

The calculation of transition dipole moments between nonorthogonal \gls*{oo} states in the PAW formalism has been implemented in the GPAW software~\cite{Mortensen2024-ji,Mortensen2005-nw}. The smooth contribution $\expv{\tilde{\psi}_i^k}{\mathbf r}{\tilde{\psi}_j^{k^\prime}}$ is evaluated on a coarse grid in the representation employed in the calculation, such as PWs or an \gls*{lcao} basis set. The PAW corrections are evaluated efficiently on atom-centered radial grids within the augmentation spheres using the partial waves and pseudo partial waves provided by the atomic setups.

\subsection{Computational Settings}
Spin-unrestricted \gls*{oo} density functional calculations of the ground and excited states of water (H$_2$O), ammonia (NH$_3$), formaldehyde (CH$_2$O), methanol (CH$_3$OH), and ethylene (C$_2$H$_4$) are carried out using the GPAW~\cite{Mortensen2024-ji,Mortensen2005-nw} software v.~25.7.1b1. The calculations are done with the xc functionals PBE~\cite{Perdew1997-rt,Perdew1996-il}, PBE0~\cite{Perdew1996-pr}, and PBE with
explicit \gls*{sic}. The \gls*{sic} calculations are carried out using the Perdew–Zunger scheme~\cite{Perdew1981-oz} outlined in Section~\ref{sec:sic}, in combination with complex-valued orbitals, employing two global scaling factors: $\alpha=1$, denoted as PBE-SIC, and $\alpha=1/2$, denoted as PBE-SIC/2. All calculations use the PAW approach~\cite{Blochl1994-gk} with the frozen-core approximation for the orbitals. Earlier benchmark studies by Loos and co-workers~\cite{Loos2018-ww} report that frozen-core effects on the excitation energy for similarly sized small molecules are negligible for post-Hartree-Fock methods, typically on the order of 0.01~eV. \gls*{lcao} and  PW basis set representations are used. \Gls*{pw} calculations use a kinetic-energy cutoff of 1200~eV. Convergence with respect to this cutoff was assessed for methanol, which includes the most diffuse excited states in the present
benchmark. Lowering the cutoff to 1000~eV while keeping the simulation box
and real-space grid spacing unchanged alters the excitation energies and
permanent dipole moment components by approximately $10^{-3}$~eV and
$10^{-3}$~D, respectively, and the oscillator strengths by less than
$10^{-5}$. The PW results are taken as a basis set reference because they
are typically consistent with results obtained using a real-space-grid
representation, as also found for excitation energy and permanent dipole
moments in the accompanying article~\cite{Restaino2026-iv}. In the \gls*{lcao} calculations, the valence electrons are represented with atomic basis functions built from primitive Gaussian-type functions taken from the aug-cc-pVDZ~\cite{Pritchard2019-vq,Kendall1992-wv,Dunning1989-vi} or d-aug-cc-pVDZ~\cite{Pritchard2019-vq,Woon1994-fz,Kendall1992-wv,Dunning1989-vi} basis sets. In both cases, the Gaussian basis is supplemented by one set of ``sz'' numerical atomic orbitals~\cite{Rossi2015-za}. In this hybrid strategy, the numerical atomic orbitals provide a compact description of the atomic ground-state valence space, while the Gaussian functions supply diffuse and polarization flexibility for excited states. Contracted functions of type "s" are removed from the \gls*{lcao} basis sets to avoid redundancy with the atom-centered partial waves in the augmentation spheres used in the \gls*{paw} formalism. For the same reason, double-$\zeta$ quality bases are used here, while the use of triple-$\zeta$ bases is not expected to change the
calculated properties, consistent with the generally close agreement
between the d-aug and PW results found here. All calculations are performed for isolated gas-phase molecules in $C_1$ symmetry. Each molecule is placed in a cubic simulation box with at least 10.5~{\AA} of vacuum around the molecule. The effect of the size of the simulation box was assessed for the S$_3$ and S$_4$ states of methanol, the most diffuse excited states considered in this work. As reported in Table~S1 of the \gls*{si}, increasing the vacuum around the atoms to 15~{\AA} changes the excitation energy and the oscillator strengths by less than $0.5\cdot 10^{-2}$~eV and $0.5\cdot 10^{-5}$, respectively.
 The pseudo orbitals for the valence electrons and electron density are represented on a uniform real-space grid with spacing 0.16~{\AA}.  

A direct optimization approach based on a limited-memory symmetric rank-1 (L-SR1) algorithm~\cite{Ivanov2021, Levi2020-nz} is used to converge on the OO excited-state solutions. The starting guess is taken as the ground-state orbitals, but with a nonaufbau occupation pattern that reflects the promotion of a single electron, either within the same spin channel for mixed-spin states or between different spin channels for triplet states. Throughout the SCF procedure, the character of occupied and virtual orbitals is tracked by analyzing their orbital characters using \gls*{mom}~\cite{Barca2018}. The orbitals involved in the excitations obtained in PBE calculations are shown in Figures~S1--S5 of the \gls*{si}.
All excited states are open-shell singlets, and their energy and transition dipole moment with the ground state are obtained using the approximate spin purification formulas~\cite{Ziegler1977-xv}
\begin{align}\label{eq:energy_spin_purification}
    E_{\mathrm{S}} = 2 E_{\mathrm{M}} - E_{\mathrm{T}} \\ 
    \bm{\mu} ^{0k}_{\mathrm{S}} = \sqrt{2}  \bm{\mu} ^{0k}  _{\mathrm{M}} \, ,
\end{align}
where $E_{\mathrm{M}}$ is the energy of the mixed-spin solution, $E_{\mathrm{T}}$ the energy of the corresponding triplet state, and $\bm{\mu} ^{0k}  _{\mathrm{M}}$ the TDM of the mixed-spin state (see Appendix \ref{app:spin_purification_tdm} for a derivation of the spin-purification formula for the TDM).

In some cases, the excitation involves a pair of degenerate orbitals, such as the 3p lone pair orbitals on the nitrogen atom of ammonia, which are unoccupied in the ground state. The occupation of just one of the two degenerate orbitals would break the spatial symmetry of the density~\cite{Barreiro-Lage2026, Ivanov2021, Sigurdarson2023-do}. To describe a single-electron excitation into the degenerate pair while preserving the symmetry of the density, complex $\mathrm{p}_{\pm}$ orbitals are used corresponding to the following linear combination of real $\mathrm{p}_x$ and $\mathrm{p}_y$ orbitals
\begin{equation}
    3\mathrm{p}_{\pm} = 3\mathrm{p}_x \pm i\,3\mathrm{p}_y \,.
\end{equation}

Calculations of the excited states are also performed using \gls*{lrdft} with the PBE and PBE0 functionals and the d-aug-cc-pVTZ basis set. Additionally, for methanol, reference calculations are performed using equation-of-motion coupled-cluster singles and doubles (EOM-CCSD)~\cite{Krylov2008-nk,Kowalski2000-gw,Nooijen1995-zu,Stanton1993-ze,Geertsen1989-ib,Rowe1968-hu} with the same basis set. These calculations are carried out with ORCA v.~6.1.0~\cite{Neese2025-cc,Neese2012-bk}. To evaluate the basis set convergence for the \gls*{lrdft} calculations, we additionally compute the excited states of methanol with the d-aug-cc-pVQZ basis set using the PBE and PBE0 functionals. For all states considered, the deviation relative to the triple-$\zeta$ is less than $10^{-2}$~eV in the excitation energy
and $3\times10^{-4}$ in the oscillator strength, as reported in the SI (Tables~S2 and S3).

Reference values of excitation energy and oscillator strength for water are taken from multistate complete active space second-order perturbation theory (MS-CASPT2) calculations with the ANO-L+R basis set~\cite{Rubio2008-lk}. For ammonia, CCSDT/d-aug-cc-pVQZ reference values were obtained through personal correspondence with the authors of Ref.~\cite{Chrayteh2021-ib}. For formaldehyde and ethylene, we use the theoretical best estimates reported by Chrayteh et al.~\cite{Chrayteh2021-ib}, which combine high-order coupled-cluster corrections with basis-set extrapolations toward the complete basis set limit.

The ground-state geometries of water, formaldehyde, ammonia, and ethylene are taken from ref.~\cite{Loos2018-ww} where they were computed at the CC3/aug-cc-pVTZ level of theory, while the geometry of methanol is taken from ref.~\cite{Lange2020-sy} where it was optimized at the CCSD(T)/aug-cc-pVQZ level of theory.

Oscillator strengths for transitions from the ground state to the excited states are evaluated using eq\ \eqref{eq:osc_strength}. For the \gls*{oo} calculations, the spin-purified excitation energy and \gls*{tdm} are used. The absorption spectra are then constructed as sums of Gaussian line shapes centered at the calculated vertical excitation energy and weighted by the corresponding oscillator strength, using a \gls*{fwhm} of 0.4~eV.

\section{Results}
\subsection{Basis set effect}
Table~\ref{tab:basis_table} compares the vertical excitation energy and oscillator strength of singlet excited states of water, formaldehyde, ammonia, methanol, and ethylene obtained in OO calculations with the PBE functional using the LCAO \gls*{aug}, \gls*{daug} basis sets as well as the \gls*{pw} representation.
\begin{table}[htb!]
\centering
\caption{%
Spin-purified vertical excitation energy, $\Delta E$ (eV), and oscillator strength, $f$, of singlet excited states of water, formaldehyde, ammonia, methanol, and ethylene computed with orbital-optimized calculations with the PBE functional and several basis sets. For the plane-wave representation, the variance of the position operator, $\sigma$ (bohr$^2$), is also reported for the ground and excited states. The excited states are identified as Rydberg (R) or valence (V) states. Transitions forbidden by symmetry are indicated by ``Forb.''
}
\label{tab:basis_table}
\begin{tabular}{llccccccc}
\hline
 &  & \multicolumn{2}{c}{aug-cc-pVDZ+sz} & \multicolumn{2}{c}{d-aug-cc-pVDZ+sz} & \multicolumn{3}{c}{plane waves} \\
\cmidrule(lr){3-4}
\cmidrule(lr){5-6}
\cmidrule(lr){7-9}
State & Character & $\Delta E$ & $f$ & $\Delta E$ & $f$ & $\Delta E$ & $\sigma$ & $f$ \\
\hline

\multicolumn{9}{l}{H$_2$O} \\
S$_0$ $A_1$ &  &  &  &  &  &  & 13.3 &  \\
S$_1$ $B_1$ & 2p$_x \rightarrow$ 3s (V/R) & 7.47 & 0.051 & 7.47 & 0.047 & 7.44 & 26.6 & 0.047 \\
S$_2$ $A_2$ & 2p$_x \rightarrow$ 3p$_y$ (R) & 8.95 & Forb. & 8.94 & Forb. & 8.90 & 37.0 & Forb. \\
S$_3$ $A_1$ & 2p$_z \rightarrow$ 3s (R) & 9.79 & 0.139 & 9.80 & 0.140 & 9.73 & 28.4 & 0.140 \\
S$_4$ $A_1$ & 2p$_x \rightarrow$ 3p$_x$ (R) & 11.24 & 0.000 & 9.90 & 0.002 & 9.84 & 57.9 & 0.003 \\
S$_5$ $B_1$ & 2p$_x \rightarrow$ 3p$_z$ (R) & 11.55 & 0.004 & 9.90 & 0.008 & 9.83 & 57.8 & 0.010 \\[0.6em]

\multicolumn{9}{l}{CH$_2$O} \\
S$_0$ $A_1$ &  &  &  &  &  &  & 23.4 &  \\
S$_1$ $A_2$ & 2p$_y \rightarrow \pi^*$ (V) & 3.58 & Forb. & 3.58 & Forb. & 3.54 & 24.1 & Forb. \\
S$_2$ $B_2$ & 2p$_y \rightarrow$ 3s (R) & 6.91 & 0.066 & 6.90 & 0.065 & 6.88 & 52.5 & 0.065 \\
S$_3$ $B_2$ & 2p$_y \rightarrow$ 3p$_z$ (R) & 7.75 & 0.031 & 7.62 & 0.026 & 7.59 & 62.4 & 0.027 \\
S$_4$ $A_1$ & 2p$_y \rightarrow$ 3p$_y$ (R) & 7.72 & 0.085 & 7.68 & 0.081 & 7.68 & 67.8 & 0.079 \\
S$_5$ $A_1$ & $\pi \rightarrow \pi^*$ (V) & 8.57 & 0.519 & 8.57 & 0.520 & 8.56 & 27.9 & 0.518 \\[0.6em]

\multicolumn{9}{l}{NH$_3$} \\
S$_0$ $A_1$ &  &  &  &  &  &  & 16.9 &  \\
S$_1$ $A_1$ & 2p$_z \rightarrow$ 3s (R) & 6.45 & 0.093 & 6.45 & 0.089 & 6.42 & 37.2 & 0.089 \\
S$_2$ $E$   & 2p$_z \rightarrow$ 3p$_{+}$ (R) & 7.82 & 0.015 & 7.80 & 0.016 & 7.83 & 54.7 & 0.016 \\
S$_3$ $A_1$ & 2p$_z \rightarrow$ 3p$_z$ (R) & 9.20 & 0.002 & 8.31 & 0.000 & 8.26 & 71.6 & 0.000 \\[0.6em]

\multicolumn{9}{l}{CH$_3$OH} \\
S$_0$ $A^{\prime}$ &  &  &  &  &  &  & 27.6 &  \\
S$_1$ $A^{\prime\prime}$ & 2p$_y \rightarrow$ 3s (R) & 6.36 & 0.004 & 6.35 & 0.004 & 6.34 & 45.5 & 0.004 \\
S$_2$ $A^{\prime\prime}$ & 2p$_y \rightarrow$ 3p$_z$ (R) & 7.40 & 0.042 & 7.38 & 0.039 & 7.38 & 59.8 & 0.040 \\
S$_3$ $A^{\prime\prime}$ & 2p$_y \rightarrow$ 3p$_x$ (R) & 7.90 & 0.002 & 7.74 & 0.002 & 7.80 & 76.7 & 0.002 \\
S$_4$ $A^{\prime}$       & 2p$_y \rightarrow$ 3p$_y$ (R) & 8.06 & 0.046 & 7.90 & 0.028 & 7.89 & 83.0 & 0.027 \\
S$_5$ $A^{\prime}$       & 2p$_x$/$\sigma \rightarrow$ 3s (R) & 7.94 & 0.001 & 7.93 & 0.001 & 7.89 & 47.3 & 0.001 \\[0.6em]

\multicolumn{9}{l}{C$_2$H$_4$} \\
S$_0$ $A_g$ &  &  &  &  &  &  & 30.4 &  \\
S$_1$ $B_{1u}$ & $\pi \rightarrow \pi ^*$ (V) & 6.71 & 0.662 & 6.72 & 0.660 & 6.73 & 40.5 & 0.662 \\
S$_2$ $B_{3u}$ & $\pi \rightarrow 3$s (R) & 7.17 & 0.075 & 7.14 & 0.063 & 7.13 & 66.1 & 0.062 \\
\hline
\end{tabular}
\end{table}
Valence excitations predominantly redistribute electron density within the
molecular valence region, whereas Rydberg excitations populate spatially
diffuse orbitals extending beyond this region. We characterize this
distinction using the variance of the electronic position operator,
$\sigma(\mathbf r)=\ev{\mathbf r^2}-\ev{\mathbf r}^2$, relative to its ground-state
value. Valence excitations have a
variance close to that of the ground state, whereas Rydberg
excitations exhibit a significantly larger variance. The assignments
reported in Table~\ref{tab:basis_table} combine this measure of spatial
extent with the character of the orbitals involved.

For water, the oscillator strength of the lower Rydberg states is only weakly affected by the basis set. The S$_1$ (2p$_x \rightarrow$ 3s) value changes slightly from 0.051 with \gls*{aug} to 0.047 with \gls*{daug} and \glspl*{pw}, while S$_3$ (2p$_z \rightarrow$ 3s) remains essentially unchanged at 0.139--0.140. Larger changes appear for the S$_4$ (2p$_x \rightarrow$ 3p$_x$) and S$_5$ (2p$_x \rightarrow$ 3p$_z$) states, which are found to have a variance of the electronic position operator 1.5 to 2 times larger than that of the three lowest excited states in the accompanying article ref.~\citen{Restaino2026-iv} and are hence much more diffuse. For these states, \gls*{aug} overestimates the excitation energy by more than 1~eV compared with \glspl*{pw}, as also previously observed~\cite{Sigurdarson2023-do}. 
The oscillator strengths remain small, changing from 0.000 (aug) to 0.002 (d-aug), and 0.003 (PWs) for S$_4$, and from 0.004 to 0.008 and 0.010 for S$_5$, respectively. The corresponding absolute changes from aug to PWs are 0.003 and 0.006, although the relative changes are considerably larger because both transitions are weak. We find the electric dipole moment  to exhibit an even larger dependence on the basis set for these states in ref.~\citen{Restaino2026-iv}, with d-aug results still deviating significantly from the PW results.

Ammonia shows a similar trend. The oscillator strengths of the least diffuse S$_1$ (2p$_z \rightarrow$ 3s) and S$_2$ (2p$z \rightarrow$ 3p$_+$) states are nearly converged already with \gls*{aug}, changing only from 0.093 to 0.089 and from 0.015 to 0.016, respectively. By contrast, S$_3$ (2p$_z \rightarrow$ 3p$_z$), which is considerably more diffuse according to the variance listed in Table~\ref{tab:basis_table}, shows a large basis set effect in the excitation energy. The $\Delta E$ decreases from 9.20~eV with \gls*{aug} to 8.26~eV with \glspl*{pw}; while the oscillator strength remains very small, changing from 0.002 with \gls*{aug} to 0.000 with both \gls*{daug} and \glspl*{pw}.

For formaldehyde, the oscillator strengths are generally only moderately affected by the basis set representation, consistent with the less diffuse character of its excited states. The bright valence S$_5$ (\mbox{$\pi \rightarrow \pi^*$}) transition is essentially unchanged, with $f=0.518$--0.520. The Rydberg states show only moderate variations. The oscillator strength of S$_2$ (\mbox{2p$_y \rightarrow$ 3s}) remains at 0.065--0.066, while the one of S$_3$ (\mbox{2p$_y \rightarrow$ 3p$_z$}) decreases from 0.031 with aug to 0.026 and 0.027 with d-aug and PWs, respectively. The oscillator strength of S$_4$ (\mbox{2p$_y \rightarrow$ 3p$_y$}) decreases from 0.085 to 0.081 and 0.079.

The oscillator strengths of the excited states of methanol are in general weakly affected by the basis set representation. The S$_1$ (2p$_y \rightarrow$ 3s), S$_3$ (2p$_y \rightarrow$ 3p$_x$), and S$_5$ (2p$_x$/$\sigma \rightarrow$ 3s) transitions remain nearly unchanged. The largest absolute change is found for S$_4$ (2p$_y \rightarrow$ 3p$_y$), for which $f$ decreases from 0.046 with \gls*{aug} to 0.028 with \gls*{daug} and 0.027 with \glspl*{pw}, corresponding to an absolute change of 0.019 between aug and PWs. This state is the one for which we obtain the largest variance, as reported in Table~\ref{tab:basis_table}. A smaller decrease is observed for S$_2$ (2p$_y \rightarrow$ 3p$_z$), from 0.042 with aug to 0.039 and 0.040 for d-aug and PWs, respectively.

Finally, for ethylene, the bright valence S$_1$ ($\pi \rightarrow \pi^* $) transition is, as is expected, insensitive to the basis set. The Rydberg S$_2$ ($\pi \rightarrow$ 3s) state, which is only moderately diffuse, shows larger changes. The excitation energy varies from 7.17 to 7.13~eV, whereas the oscillator strength decreases from 0.075 with \gls*{aug} to 0.063 with \gls*{daug} and 0.062 with \glspl*{pw}.

Overall, the \gls*{daug} and \gls*{pw} oscillator strengths are generally close. The largest basis set effects are observed when comparing \gls*{aug} with the more diffuse representations, especially for weak transitions involving highly diffuse Rydberg states. For weak transitions, the relative changes can be large even when the corresponding absolute changes are small; however, several diffuse states also exhibit some of the largest absolute changes in oscillator strength. Bright valence transitions are essentially unaffected. In the accompanying article to this work~\cite{Restaino2026-iv}, we assess the performance of \gls*{oo} density functional calculations for the permanent dipole moment for the same set of molecules. There, we find that the dipole moments are not fully converged at the \gls*{daug} level for the diffuse Rydberg states, with differences with respect to the \glspl*{pw} representation that are later compared to the differences found here for the oscillator strength.

\subsection{Assessment of the Exchange--Correlation Functionals}
\begin{figure}[hbt]
    \centering
    \includegraphics[width=0.5\linewidth]{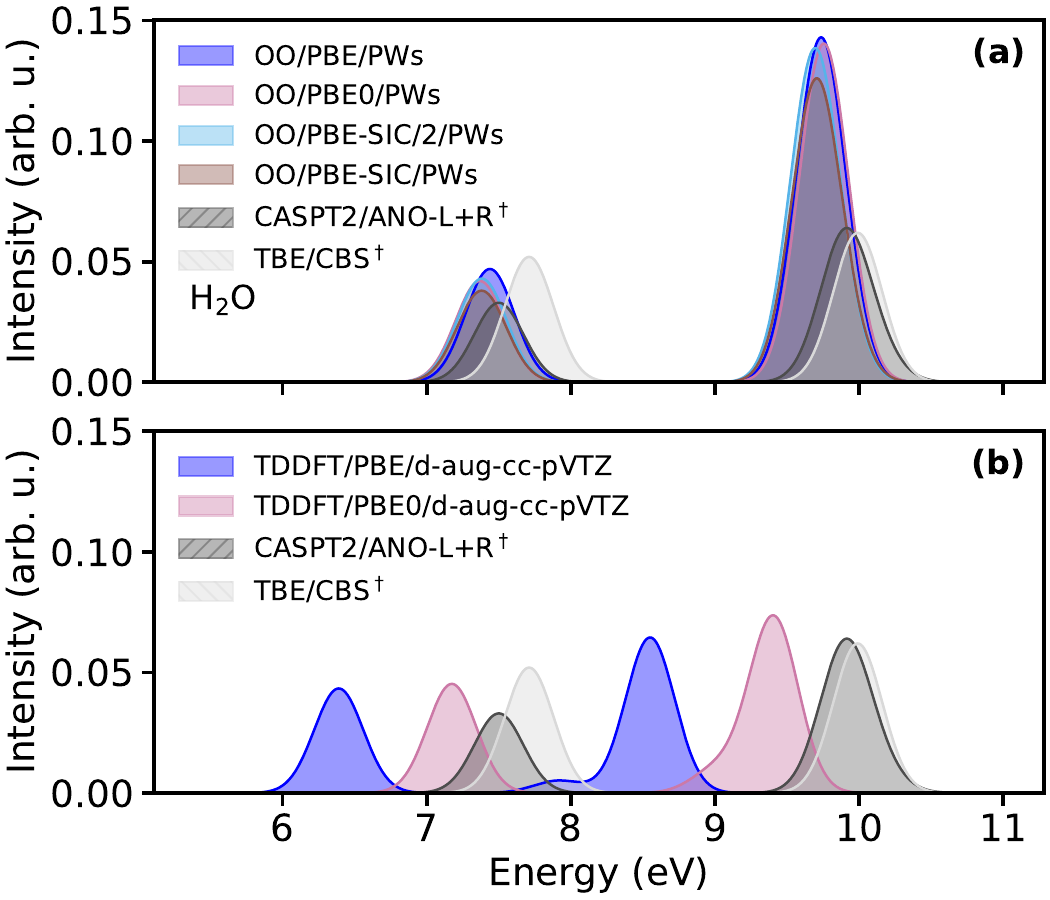}
    \caption{Calculated absorption spectrum of water. (a) Orbital-optimized (OO) density-functional and (b) linear-response time-dependent density functional theory spectra obtained using various exchange-correlation functionals, compared with results from high-level wave function-based calculations with atomic basis sets including diffuse functions, shown in gray. A Gaussian broadening of 0.4~eV \gls*{fwhm} was used to generate the spectra. The OO calculations provide an accurate estimate of the peak positions, but significantly overestimate the intensity of the S$_3$ (2p$_z \rightarrow$ 3s) transition around 9.9 eV because they do not capture the multi-configurational character of this state. $^\ddagger$MS-CASPT2/ANO-L+R values from ref.~\citen{Rubio2008-lk}; $^\ddagger$Theoretical best estimate (TBE)/complete basis set (CBS) values from ref.~\citen{Chrayteh2021-ib} obtained from high-order coupled cluster calculations and extrapolation to the complete basis set limit.}
    \label{fig:water_spectra}
\end{figure}

\begin{figure}[hbt]
    \centering
    \includegraphics[width=0.5\linewidth]{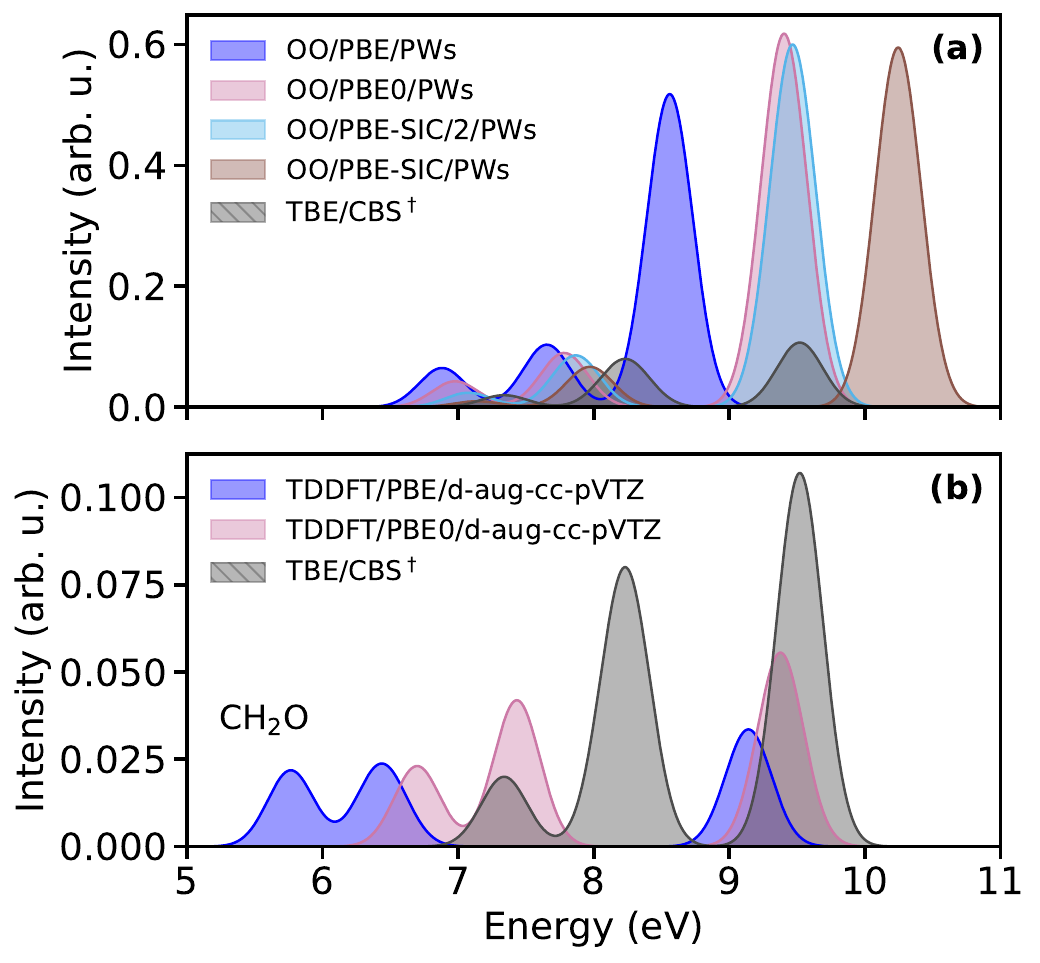}
    \caption{Calculated absorption spectrum of formaldehyde. (a) Orbital-optimized (OO) density-functional and (b) linear-response time-dependent density functional theory spectra obtained using various exchange-correlation functionals, compared with results from coupled-cluster calculations with atomic basis sets including diffuse functions, shown in gray. The two panels use different scales on the $y$-axis. A Gaussian broadening of 0.4~eV \gls*{fwhm} was used to generate the spectra. $^\dagger$Theoretical best estimate (TBE)/complete basis set (CBS) values from ref.~\citen{Chrayteh2021-ib} obtained from high-order coupled cluster calculations and extrapolation to the complete basis set limit. The OO calculations reproduce the low-energy part of the spectrum well, where Rydberg excitations appear, but significantly overestimate the intensity of the higher-energy S$_5$ ($\pi \rightarrow \pi^*$) transition, which has a multi-configurational character in higher-level calculations.}
    \label{fig:form_spectra}
\end{figure}

\begin{figure}[hbt]
    \centering
    \includegraphics[width=0.5\linewidth]{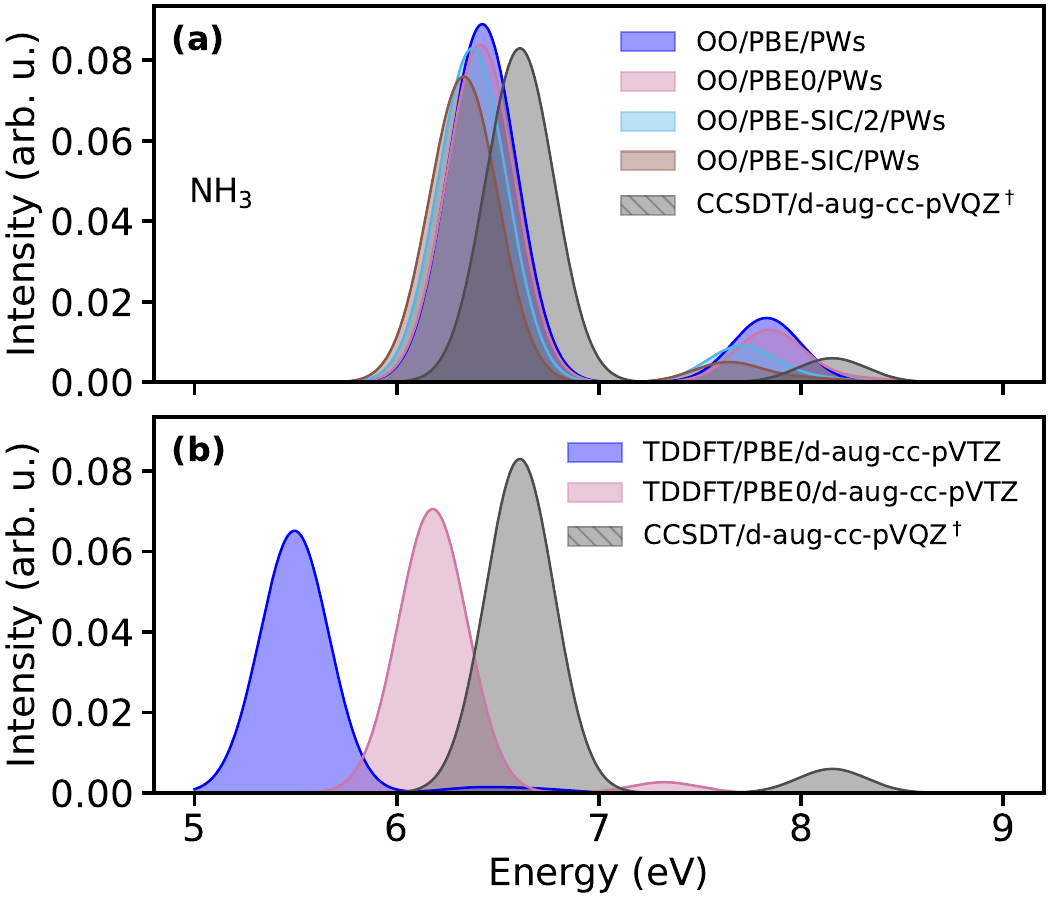}
    \caption{Calculated absorption spectrum of ammonia. (a) Orbital-optimized (OO) density-functional and (b) linear-response time-dependent density functional theory spectra obtained using various exchange-correlation functionals, compared with results from coupled-cluster calculations with atomic basis sets including diffuse functions, shown in gray. A Gaussian broadening of 0.4~eV \gls*{fwhm} was used to generate the spectra. $^\dagger$CCSDT/\mbox{d-aug-cc-pVQZ} values from ref.~\citen{perscorr_jacquemin}. The OO calculations are in remarkably good agreement with the coupled-cluster results, which is consistent with the predominantly single-configurational character of the excited states contributing to the absorption spectrum.}
    \label{fig:ammonia_spectra}
\end{figure}

\begin{figure}[hbt]
    \centering
    \includegraphics[width=0.5\linewidth]{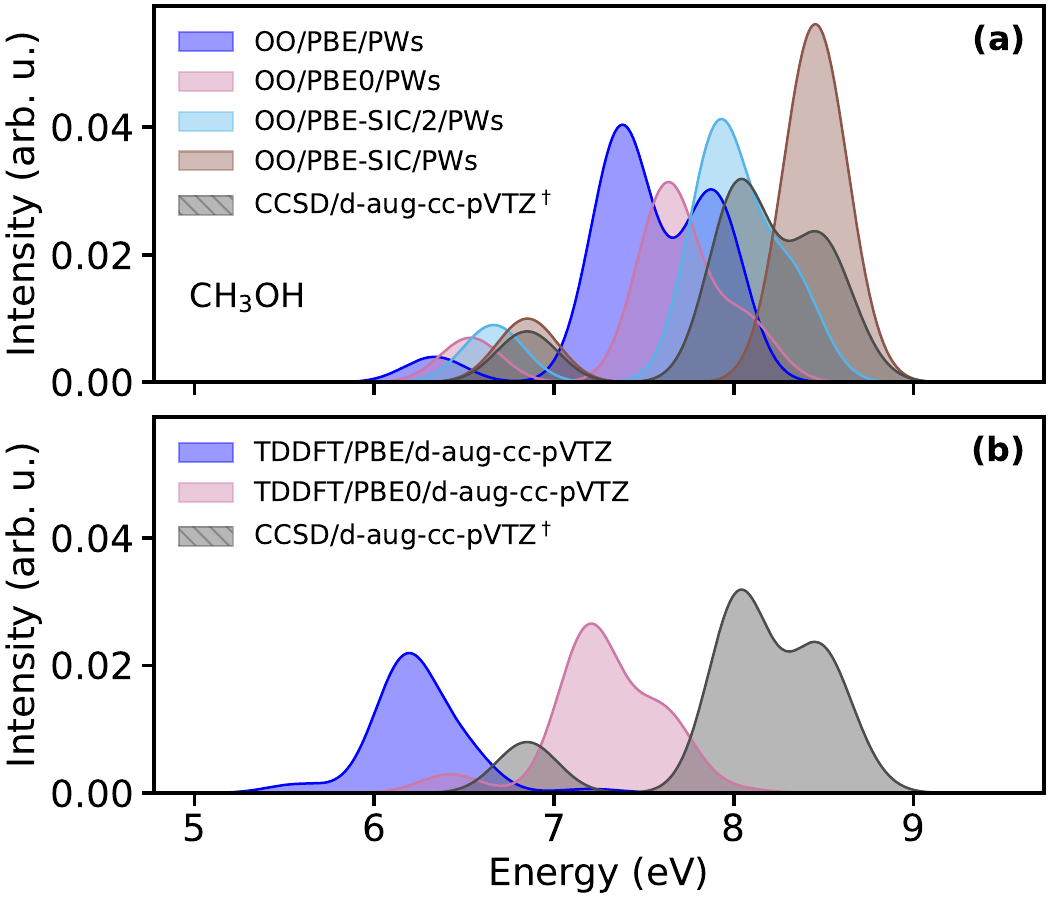}
    \caption{Calculated absorption spectrum of methanol. (a) Orbital-optimized (OO) density-functional and (b) linear-response time-dependent density functional theory spectra obtained using various exchange-correlation functionals, compared with results from coupled-cluster calculations with atomic basis sets including diffuse functions, shown in gray. A Gaussian broadening of 0.4~eV \gls*{fwhm} was used to generate the spectra. The OO calculations reproduce the main features of the coupled-cluster spectrum, with PBE-SIC/2 giving the best results in terms of shape and position of the peaks. $^\dagger$CCSD/d-aug-cc-pVTZ values from the present work.}
    \label{fig:methanol_spectra}
\end{figure}

\begin{figure}[hbt]
    \centering
    \includegraphics[width=0.5\linewidth]{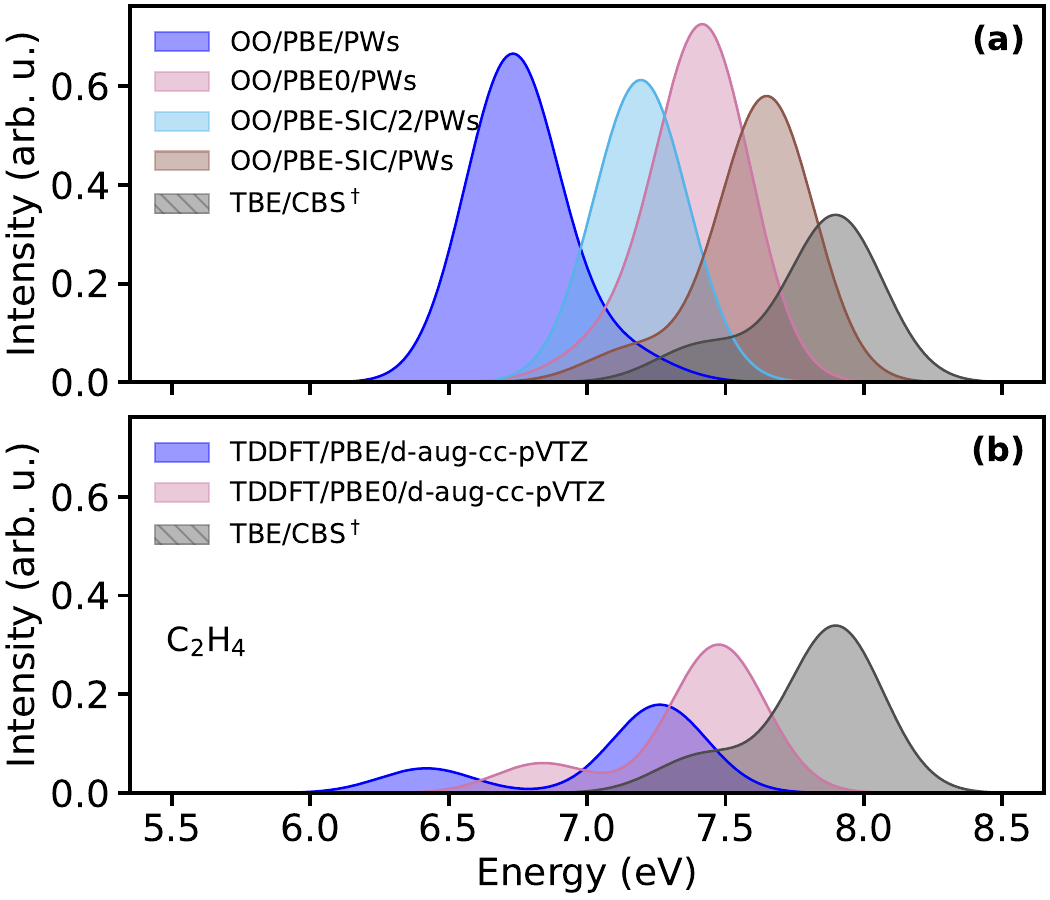}
    \caption{Calculated absorption spectrum of ethylene. (a) Orbital-optimized (OO) density-functional and (b) linear-response time-dependent density functional theory spectra obtained using various exchange-correlation functionals, compared with results from coupled-cluster calculations with atomic basis sets including diffuse functions, shown in gray. A Gaussian broadening of 0.4~eV \gls*{fwhm} was used to generate the spectra. The OO calculations significantly overestimate the intensity of the bright $\pi \rightarrow \pi^*$ transition, which has a multi-configurational character in higher-level calculations, while they reproduce the energy and intensity of the Rydberg 3s excitation much better when the PBE0, PBE-SIC/2, and PBE-SIC functionals are used. $^\dagger$Theoretical best estimate (TBE)/complete basis set (CBS) values from ref.~\citen{Chrayteh2021-ib} obtained from high-order coupled cluster calculations and extrapolation to the complete basis set limit.}
    \label{fig:ethylene_spectra}
\end{figure}

Table~\ref{tab:xc_table} compares the values of vertical excitation energy and oscillator strength obtained from \gls*{oo} calculations employing PBE, PBE0, PBE-SIC/2, and PBE-SIC with \glspl*{pw} results from high-level wave function-based calculations. The corresponding spectra are shown in Figures~\ref{fig:water_spectra}--\ref{fig:ethylene_spectra}, together with spectra computed with \gls*{lrdft}.

Table~\ref{tab:xc_table} also reports the configurational character of each excited state. States dominated by one excitation configuration are classified as predominantly single-configurational (SC), whereas states with substantial contributions from several configurations are classified as multi-configurational (MC). This assignment is based on the reported relative weights of configurations in correlated wave functions from high-level calculations in the literature.

\begin{table}[htb!]
\centering
\caption{%
Vertical excitation energy, $\Delta E$ (eV), and oscillator strength, $f$, of singlet excited states of water, formaldehyde, ammonia, methanol, and ethylene, obtained from orbital-optimized density functional calculations using various exchange-correlation functionals and a plane-wave representation of the orbitals, alongside results from high-level wave function-based calculations using atomic-orbital basis sets with diffuse functions. All OO oscillator strengths use spin-purified transition dipole moments. The OO excitation energies are also spin purified, except for entries marked $^\mathparagraph$. States are classified as predominantly single-configurational (SC) or multi-configurational (MC) based on higher-level calculations reported in the literature (see  table footnotes). Transitions forbidden by symmetry are indicated by ``Forb.''.
}
\label{tab:xc_table}
\small
\begin{tabular}{
l
c
S[table-format=2.2]
S[table-format=1.3]
S[table-format=2.2]
S[table-format=1.3]
S[table-format=2.2]
S[table-format=1.3]
S[table-format=2.2]
S[table-format=1.3]
S[table-format=2.2]
S[table-format=1.3]
}
\hline
 & & \multicolumn{2}{c}{PBE} & \multicolumn{2}{c}{PBE0} & \multicolumn{2}{c}{PBE-SIC/2} & \multicolumn{2}{c}{PBE-SIC} & \multicolumn{2}{c}{Reference} \\
\cmidrule(lr){3-4}
\cmidrule(lr){5-6}
\cmidrule(lr){7-8}
\cmidrule(lr){9-10}
\cmidrule(lr){11-12}
State & Config. & {$\Delta E$} & {$f$} & {$\Delta E$} & {$f$} & {$\Delta E$} & {$f$} & {$\Delta E$} & {$f$} & {$\Delta E$} & {$f$} \\
\hline

\multicolumn{12}{l}{H$_2$O} \\
S$_1$ $B_1$ & SC & 7.44 & 0.047 & 7.37 & 0.042 & 7.38 & 0.043 & 7.38 & 0.038 & \tablenum{7.50}$^\dagger$ & \tablenum{0.033}$^\dagger$ \\
S$_2$ $A_2$ & SC & 8.90 & \multicolumn{1}{c}{Forb.} & 8.89 & \multicolumn{1}{c}{Forb.} & 9.03 & \multicolumn{1}{c}{Forb.} & 9.15 & \multicolumn{1}{c}{Forb.} & \tablenum{9.27}$^\dagger$ & \multicolumn{1}{c}{Forb.} \\
S$_3$ $A_1$ & MC & 9.73 & 0.140 & 9.76 & 0.127 & 9.69 & 0.128 & 9.70 & 0.117 & \tablenum{9.86}$^\dagger$ & \tablenum{0.032}$^\dagger$ \\
S$_4$ $A_1$ & MC & 9.84 & 0.003 & 9.82 & 0.004 & 9.88 & 0.005 & 9.89 & 0.006 & \tablenum{9.95}$^\dagger$ & \tablenum{0.033}$^\dagger$ \\
S$_5$ $B_1$ & SC & 9.83 & 0.010 & 9.72 & 0.010 & 9.78 & 0.009 & 9.80 & 0.007 & \tablenum{10.15}$^\dagger$ & \tablenum{0.006}$^\dagger$ \\[0.6em]

\multicolumn{12}{l}{CH$_2$O} \\
S$_1$ $A_2$ & SC & 3.54 & \multicolumn{1}{c}{Forb.} & 3.46 & \multicolumn{1}{c}{Forb.} & 3.38 & \multicolumn{1}{c}{Forb.} & 3.25 & \multicolumn{1}{c}{Forb.} & \tablenum{3.99}$^\ddagger$ & \multicolumn{1}{c}{Forb.} \\
S$_2$ $B_2$ & SC & 6.88 & 0.065 & 6.98 & 0.043 & 7.08 & 0.024 & 7.14 & 0.010 & \tablenum{7.34}$^\ddagger$ & \tablenum{0.020}$^\ddagger$ \\
S$_3$ $B_2$ & SC & 7.59 & 0.027 & 7.73 & 0.028 & 7.82 & 0.032 & 7.94 & 0.029 & \tablenum{8.16}$^\ddagger$ & \tablenum{0.035}$^\ddagger$ \\
S$_4$ $A_1$ & SC & 7.68 & 0.079 & 7.81 & 0.064 & 7.90 & 0.056 & 8.00 & 0.039 & \tablenum{8.28}$^\ddagger$ & \tablenum{0.050}$^\ddagger$ \\
S$_5$ $A_1$ & MC & 8.56 & 0.518 & 9.40 & 0.618 & 9.47 & 0.600 & 10.25 & 0.595 & \tablenum{9.52}$^\ddagger$ & \tablenum{0.107}$^\ddagger$ \\[0.6em]

\multicolumn{12}{l}{NH$_3$} \\
S$_1$ $A_1$ & SC & 6.42 & 0.089 & 6.41 & 0.084 & 6.37 & 0.083 & 6.33 & 0.076 & \tablenum{6.61}$^*$ & \tablenum{0.083}$^*$ \\
S$_2$ $E$ & SC & 7.83 & 0.016 & 7.84 & 0.013
& \tablenum{7.70}\rlap{$^\mathparagraph$} & \tablenum{0.009}\rlap{$^\|$}
& \tablenum{7.64}\rlap{$^\mathparagraph$} & \tablenum{0.005}\rlap{$^\|$}
& \tablenum{8.15}\rlap{$^*$} & \tablenum{0.006}\rlap{$^*$} \\
S$_3$ $^1A_1$ & SC & 8.26 & 0.000 & 8.28 & 0.001
& \tablenum{8.08}\rlap{$^\mathparagraph$} & \tablenum{0.001}\rlap{$^\|$}
& \tablenum{8.05}\rlap{$^\mathparagraph$} & \tablenum{0.001}\rlap{$^\|$}
& \tablenum{8.60}\rlap{$^*$} & \tablenum{0.000}\rlap{$^*$} \\[0.6em]

\multicolumn{12}{l}{CH$_3$OH} \\
S$_1$ $A^{\prime\prime}$ & SC & 6.34 & 0.004 & 6.54 & 0.007 & 6.66 & 0.009 & 6.85 & 0.010 & \tablenum{6.85}$^\S$ & \tablenum{0.008}$^\S$ \\
S$_2$ $A^{\prime\prime}$ & SC & 7.38 & 0.040 & 7.63 & 0.031 & 7.92 & 0.040 & 8.42 & 0.035 & \tablenum{8.03}$^\S$ & \tablenum{0.031}$^\S$ \\
S$_3$ $A^{\prime\prime}$ & SC & 7.80 & 0.002 & 8.04 & 0.000 & 8.23 & 0.000
& \tablenum{8.41}\rlap{$^\mathparagraph$} & \tablenum{0.002}\rlap{$^\|$}
& \tablenum{8.46}\rlap{$^\S$} & \tablenum{0.001}\rlap{$^\S$} \\
S$_4$ $A^{\prime}$ & SC & 7.89 & 0.027
& \tablenum{8.06}\rlap{$^\mathparagraph$} & \tablenum{0.010}\rlap{$^\|$}
& 8.31 & 0.015 & 8.50 & 0.014
& \tablenum{8.46}\rlap{$^\S$} & \tablenum{0.019}\rlap{$^\S$} \\
S$_5$ $A^{\prime}$ & SC & 7.89 & 0.001 & 8.21 & 0.000 & 8.34 & 0.002 & 8.56 & 0.008 & \tablenum{8.63}$^\S$ & \tablenum{0.004}$^\S$ \\[0.6em]

\multicolumn{12}{l}{C$_2$H$_4$} \\
S$_1$ $B_{1u}$ & MC & 6.73 & 0.662 & 7.42 & 0.718 & 7.20 & 0.551 & 7.65 & 0.579 & \tablenum{7.90}$^\ddagger$ & \tablenum{0.338}$^\ddagger$ \\
S$_2$ $B_{3u}$ & SC & 7.13 & 0.062 & 7.06 & 0.070 & 7.16 & 0.063 & 7.17 & 0.068 & \tablenum{7.42}$^\ddagger$ & \tablenum{0.076}$^\ddagger$ \\
\hline

\multicolumn{12}{l}{\footnotesize $^\mathparagraph$Excitation energy of mixed-spin solution, the corresponding triplet calculation did not converge.} \\
\multicolumn{12}{l}{\footnotesize $^\|$Oscillator strength evaluated using the mixed-spin excitation energy and a spin-purified TDM.} \\
\multicolumn{12}{l}{\footnotesize $^\dagger$MS-CASPT2/ANO-L+R values from ref.~\citen{Rubio2008-lk}; $^*$CCSDT/d-aug-cc-pVQZ values from ref.~\citen{perscorr_jacquemin};} \\
\multicolumn{12}{l}{\footnotesize $^\S$CCSD/d-aug-cc-pVTZ values from this work; $^\ddagger$TBE/CBS values from ref.~\citen{Chrayteh2021-ib}.} \\

\end{tabular}
\end{table}

Figure~\ref{fig:water_spectra} displays the absorption spectra of water obtained from \gls*{oo} and \gls*{lrdft} calculations, alongside spectra computed from high-level MS-CASPT2 and coupled-cluster calculations. At the \gls*{oo} level, the peak positions agree remarkably well with those of the higher-level multi-configurational calculations, even with the simple PBE functional. In contrast, the \gls*{lrdft} spectra are red-shifted, with a red shift as large as 1~eV with PBE. The height of the S$_1$ peak is also reproduced well by the OO calculations. The main discrepancy between the \gls*{oo} spectra and the spectra obtained in the higher-level calculations affects the intensity of the S$_3$ and S$_4$ transitions. As shown in Table~\ref{tab:xc_table}, the oscillator strength of S$_3$ is overestimated by all \gls*{oo} calculations, with values of 0.140, 0.127, 0.128, and 0.117 for PBE, PBE0, PBE-SIC/2, and PBE-SIC, respectively. These values exceed both the MS-CASPT2/ANO-L+R estimate of 0.032~\cite{Rubio2008-lk} and the CBS/TBE estimate of 0.062~\cite{Chrayteh2021-ib}. The opposite trend is found for the S$_4$ state, where the oscillator strength is underestimated by about one order of magnitude relative to the MS-CASPT2 value for all functionals. The S$_3$ state belongs to the same irreducible representation as the ground state, A$_1$. It is therefore important to assess whether the large \gls*{tdm} associated with this transition is artificially amplified by the nonzero overlap between the ground and excited states introduced by the nonorthogonality of the \gls*{oo} states. The overlaps reported in Table~S5 of the \gls*{si} show that this is unlikely to be the dominant source of the error. For S$_3$, PBE/\gls*{aug} and PBE/\gls*{daug} give overlaps of 0.0543 and 0.0516, respectively, while the PBE/\glspl*{pw} value is smaller, 0.0226. Hybrid exchange further reduces this overlap: With PBE0/\glspl*{pw}, the S$_3$ overlap falls below $10^{-5}$. Similarly, PBE-SIC/2/\glspl*{pw} yields a small S$_3$ overlap of 0.0023. Despite the significantly different values of overlap, all functionals overestimate the intensity of the S$_3$. As shown and extensively discussed below, the main source of error affecting the oscillator strength of the S$_3$ and S$_4$ states is the multi-configurational character of these states, which is not captured by OO calculations.

The interpretation of the LR-TDDFT spectrum of water in Figure~\ref{fig:water_spectra}(b) also requires some care. In the MS-CASPT2 calculations, the S$_3$ and S$_4$ states are close in energy, at 9.86 and 9.95~eV, and have the same intensity. In the spectrum, their contributions add up and produce the peak visible at higher energy. The \gls*{lrdft} PBE and PBE0 spectra exhibit a red-shifted peak of similar intensity, but this peak arises solely from the S$_3$ transition. Indeed, instead of predicting similar intensity for the S$_3$ and S$_4$ states, \gls*{lrdft} overestimates the S$_3$ intensity and places S$_4$ at lower energy than S$_3$, which gives rise to the small shoulder on the low-energy side of the peak. Thus, while the MS-CASPT2 and \gls*{lrdft} spectra appear to have a similar shape, the \gls*{lrdft} calculations do not capture the correct S$_3$/S$_4$ intensity pattern.

Figure~\ref{fig:form_spectra} shows the calculated absorption spectra of formaldehyde. The \gls*{oo} calculations reproduce the features of the lower-energy region of the spectrum quite well, where the weaker Rydberg states appear, with the best agreement with the coupled-cluster spectra obtained for the PBE0 and PBE-SIC/2 functionals. In contrast, the \gls*{lrdft} calculations place the Rydberg states too low in energy and give significantly less accurate relative intensities. PBE0 improves the values of \gls*{lrdft} excitation energy compared to PBE, but the Rydberg region remains red-shifted relative to the higher-level calculations by approximately 1~eV and the height of the second bright excitation is significantly underestimated.
The prominent feature above 9~eV in the \gls*{oo} spectrum corresponds to the transition to the S$_5$, $\pi \rightarrow \pi^*$ valence state. For this state, the oscillator strength is strongly overestimated in the \gls*{oo} calculations. The values of oscillator strength reported in Table~\ref{tab:xc_table} are 0.518, 0.618, 0.600, and 0.595 for PBE, PBE0, PBE-SIC/2, and PBE-SIC, respectively, compared to the CBS/TBE value of 0.107. While the intensity is only weakly affected by the choice of \gls*{xc} functional, the excitation energy varies more, with PBE0 and PBE-SIC/2 providing closer agreement with the CBS/TBE results. \Gls*{lrdft} gives several states with a contribution of $\pi \rightarrow \pi^*$ excitation. For the comparison in Figure~\ref{fig:form_spectra}(b), the state with the largest contribution is chosen, corresponding to a weight of the $\pi \rightarrow \pi^*$ excitation of around 38\% and 49\% with PBE and PBE0, respectively. With this assignment, \gls*{lrdft} captures the position of the peak corresponding to the $\pi \rightarrow \pi^*$ transition in the coupled-cluster spectrum better than the \gls*{oo} calculations, but its intensity is underestimated with both functionals.

The absorption spectra of ammonia are shown in Figure~\ref{fig:ammonia_spectra}. For this molecule, the \gls*{oo} calculations are in remarkably good agreement with the CCSDT results. The lowest transition has an oscillator strength of 0.089 with PBE, 0.084 with PBE0, 0.083 with PBE-SIC/2, and 0.076 with PBE-SIC, comparing well with the CCSDT value of 0.083. The second lowest excitation, with $E$ symmetry and 3p$_{+}$ Rydberg character remains weak for all \gls*{oo} functionals, with $f=0.005-0.016$, while S$_3$ is essentially dark, in agreement with the CCSDT calculations. The \gls*{lrdft} spectra in Figure~\ref{fig:ammonia_spectra}(b) are shifted to lower energy and underestimate the peak of the 3p$_{+}$ transition relative to the CCSDT spectrum.

The calculated absorption spectra of methanol are shown in Figure~\ref{fig:methanol_spectra}. The CCSD spectrum contains a small peak around 6.85 eV arising from the S$_1$ transition and two main features, between 8.0 and 8.7~eV, arising mainly form the S$_2$ and S$_4$ transitions, respectively. In the \gls*{oo} calculations, the spectrum is progressively blue shifted with PBE, PBE0, PBE-SIC/2, and PBE-SIC. PBE places the dominant intensity at too low energy, while PBE0 and PBE-SIC/2 move the spectrum closer to the CCSD spectrum. PBE-SIC shifts the peaks too high in energy. For S$_1$, the oscillator strength increases from 0.004 with PBE to 0.010 with PBE-SIC, and the closest agreement with the CCSD reference value of 0.008 is obtained with PBE-SIC/2 and PBE-SIC. For the highest intensity S$_2$ transition, PBE0 reproduces the CCSD value of oscillator strength of 0.031, whereas PBE and PBE-SIC/2 give slightly larger values of 0.040. The S$_3$ transition remains very weak for all functionals, consistent with the CCSD result. For S$_4$, PBE-SIC/2 gives an oscillator strength closer to the CCSD value of 0.019, while PBE and PBE0 give a too large and too low value, respectively. For S$_5$, the functional dependence is more pronounced, although the intensity of this transition remains low with all functionals. Overall, PBE-SIC/2 gives a shape of the spectrum and position of the peaks in closer agreement with the CCSD spectrum. The \gls*{lrdft} spectra are red-shifted relative to CCSD by about 2~eV with PBE and about 1~eV with PBE0, while the intensities are consistently underestimated.

Finally, Figure~\ref{fig:ethylene_spectra} shows the absorption spectra of ethylene. The coupled-cluster spectrum is dominated by the bright $\pi \rightarrow \pi^*$ transition around 7.9 eV and by a smaller peak corresponding to excitation to the 3s Rydberg state at around 7.4 eV. In the OO calculations, the relative position of the two transitions is functional-dependent. With PBE, the energy ordering of the two peaks is reversed compared to coupled cluster, the bright $\pi \rightarrow \pi^*$ valence transition lying below the weaker Rydberg transition.
With PBE0, PBE-SIC/2, and PBE-SIC, the \mbox{$\pi \rightarrow \pi^*$} state shifts to an excitation energy higher than the Rydberg state, with PBE-SIC predicting a gap between the two states closer to the coupled-cluster results. All \gls*{oo} functionals overestimate the intensity of the valence excitation, giving an oscillator strength of 0.662 with PBE, 0.718 with PBE0, 0.551 with PBE-SIC/2, and 0.579 with PBE-SIC, compared to the CBS/TBE reference value of 0.338. The PBE-SIC calculations reduce the intensity relative to PBE and PBE0, but the transition remains too intense. Overall, for this molecule, PBE-SIC provides a spectrum closer to the one predicted by the high-level coupled-cluster calculations. \gls*{lrdft} also struggles to describe the spectrum of ethylene. As in formaldehyde, the \gls*{lrdft} calculations with PBE predict several states with contribution of the $\pi \rightarrow \pi^*$ excitation. The state included in the spectrum shown in Figure~\ref{fig:ethylene_spectra}(b) has the largest weight, $\sim$57\%. In contrast, PBE0 produces a single state with a dominant $\pi \rightarrow \pi^*$ contribution, leading to a peak with a position and intensity closer to the CBS/TBE reference. The intensity, however, remains underestimated. The 3s Rydberg transition is systematically predicted too low in energy.

The state-by-state comparisons discussed above reveal that the largest
deviations in the OO oscillator strengths are concentrated in a small
subset of excited states rather than being characteristic of a particular
molecule or type of excitation. These include both the Rydberg S$_3$ and
S$_4$ states of water and the valence $\pi \rightarrow \pi^*$ S$_5$ and
S$_1$ states of formaldehyde and ethylene, respectively. What these states
have in common is their pronounced multi-configurational character in
higher-level wave function-based calculations. This suggests that these
large errors are associated primarily with limitations of the mean-field
OO description rather than with the particular exchange--correlation
functional employed.

\begin{figure}[hbt]
    \centering
    \includegraphics[width=\linewidth]{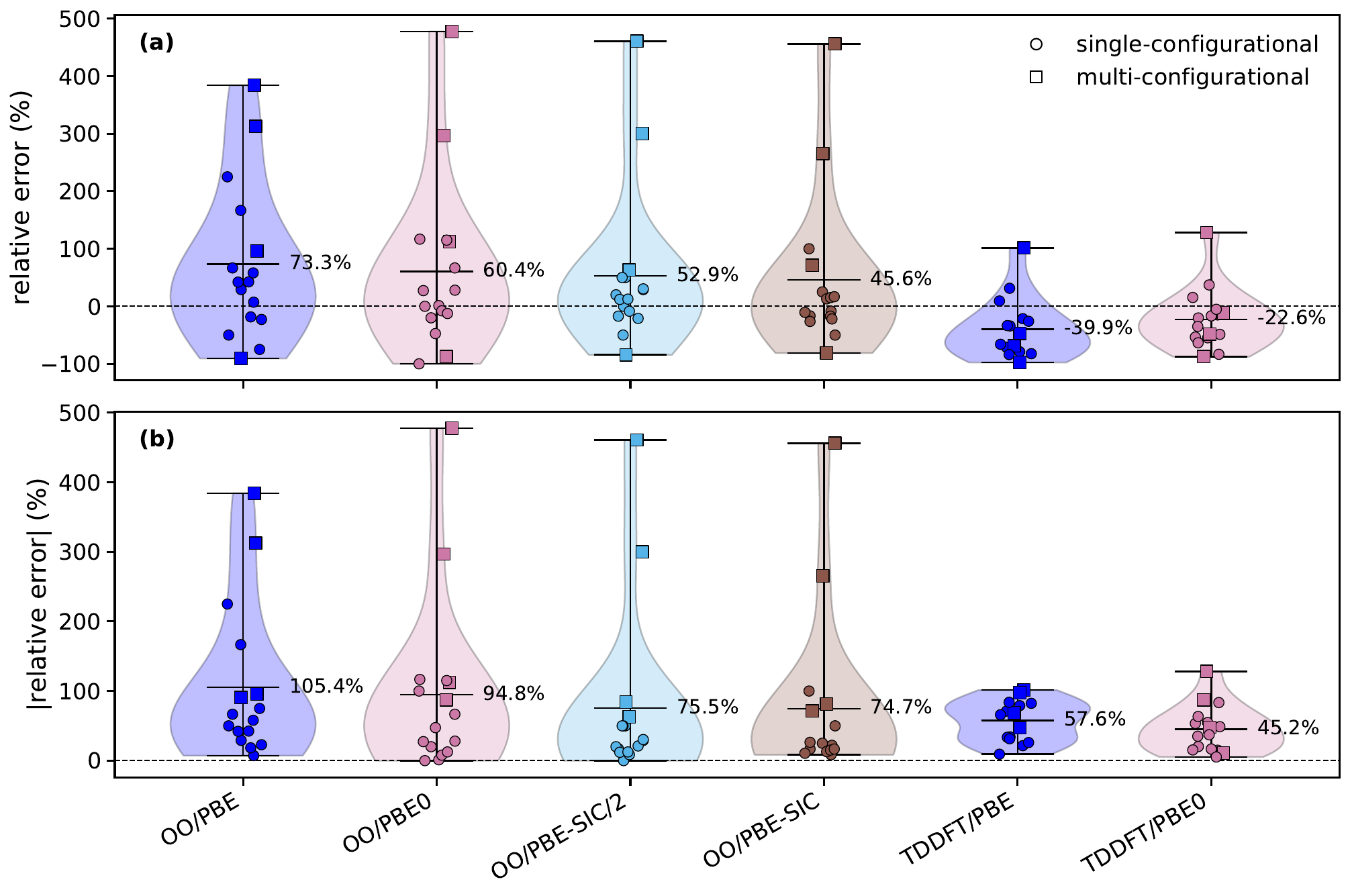}
    \caption{
    Signed (a) and absolute (b) relative errors in oscillator strength for orbital-optimized (OO) and linear-response time-dependent density functional theory (LR-TDDFT) for various exchange-correlation functional. The mean value is indicated as a horizontal bar. Only states with $f_{\mathrm{ref}} > 0.001$ are included. Excited states with multi-configurational character in higher-level calculations from the literature are marked with squares, those with single-configurational character with circles. The OO calculations provide large errors for multi-configurational states compared to LR-TDDFT, independently of the functional.}
    \label{fig:relative_error}
\end{figure}
Figure~\ref{fig:relative_error} shows the distribution of the signed and absolute relative error on the oscillator strength for each functional, defined as
\begin{equation}
\mathrm{err} = \dfrac{f_{\mathrm{OO}}-f_{\mathrm{ref}}}{f_{\mathrm{ref}}} \cdot 100\% \, ,
\end{equation}
for all states with $f_{\mathrm{ref}} > 0.001$, where the reference values are taken as the coupled-cluster results reported in Table \ref{tab:xc_table}.
The distribution of signed errors shows that all functionals employed in the \gls*{oo} calculations in most cases overestimate the reference values, although they can also yield negative deviations from the reference values. In contrast, the \gls{lrdft} distributions display almost exclusively negative deviations, indicating that they systematically underestimate the reference values of oscillator strength.
The absolute relative errors of the OO calculations show broad distributions for all \gls*{xc} functionals. Among them, PBE-SIC gives the lowest mean absolute relative error, but all functionals display outliers that significantly distort the mean. These outliers with much larger errors compared to the other states correspond to excited states that have a significant multi-configurational character in high-level correlated wave function calculations: the S$_3$ and S$_4$ Rydberg states of water as well as the S$_5$ and S$_1$ states of formaldehyde and ethylene, respectively, with a prevalent $\pi \rightarrow \pi^*$ character, identified also above in the analysis of the individual spectra. Due to these outliers, the mean errors remain large for all functionals, without a clear improvement of the hybrid and self-interaction corrected functionals over PBE. This is consistent with the molecule-by-molecule analysis, where changes in the functional improved selected transitions but did not correct the largest failures. Compared with the \gls*{oo} calculations, \gls*{lrdft} gives narrower error distributions and smaller mean absolute errors for the oscillator strengths, thanks mostly to significantly smaller errors for the multi-configurational states. 
\begin{figure}[hbt]
    \centering
    \includegraphics[width=\linewidth]{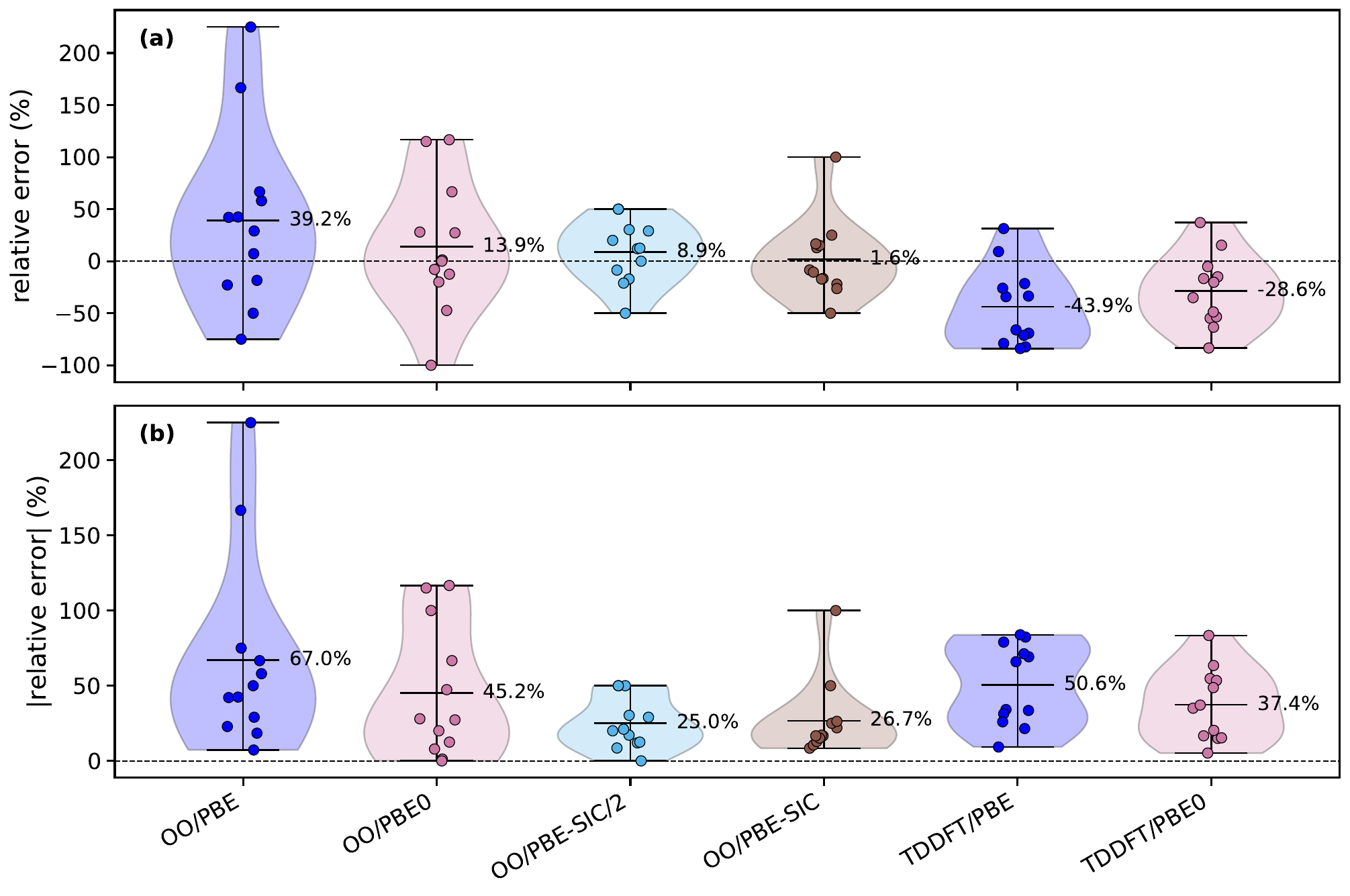}
    \caption{Signed (a) and absolute (b) relative errors in oscillator strength for orbital-optimized (OO) and linear-response time-dependent density functional theory (LR-TDDFT) for various exchange-correlation functional. The mean value is indicated as a horizontal bar. Only states with $f_{\mathrm{ref}} > 0.001$ and a single-configurational character in higher-level calculations are included. When multi-configurational states are excluded, the performance of the \gls*{oo} calculations becomes comparable to that of \gls*{lrdft}, with PBE-SIC/2 yielding the lowest mean error among the functionals considered.}
    \label{fig:relative_error_no_multi}
\end{figure}
Figure\ \ref{fig:relative_error_no_multi} shows that the mean relative errors of the OO calculations decrease significantly when the analysis is restricted to states that have single-configurational character in higher-level calculations and become more similar to the errors of LR-TDDFT. The OO PBE, PBE0, and PBE-SIC/2 mean absolute relative errors are reduced to 67\%, 45\%, and 25\%, respectively, a decrease by more than a factor of two compared to the case where the multi-configurational states are included.  The \gls*{lrdft} mean errors instead are largely unaffected by the exclusion of the multi-configurational states. An analysis of the error on the oscillator strength without percentage normalization is also reported in Figures~S7 and S8 of the \gls*{si}, showing both the signed error, $\Delta f=f-f_{\mathrm{ref}}$, and its absolute value. For the single-configurational states, the mean absolute error in oscillator strength is slightly smaller with OO/PBE than with LR-TDDFT/PBE and smaller with OO/PBE0 than with LR-TDDFT/PBE0. Including SIC further reduces the OO errors, giving smaller mean absolute errors than LR-TDDFT with either functional.

Taken together, Figures~\ref{fig:relative_error} and \ref{fig:relative_error_no_multi} therefore distinguish two different sources of error. For predominantly single-configurational states, the accuracy of the oscillator strength depends appreciably on the
xc functional, with the inclusion of exact exchange and self-interaction correction providing a significant improvement. For multi-configurational states, instead, the error associated with representing the excited state without configurational mixing dominates over the differences between the functionals considered here.

\begin{figure}[hbt]
    \centering
    \includegraphics[width=\linewidth]{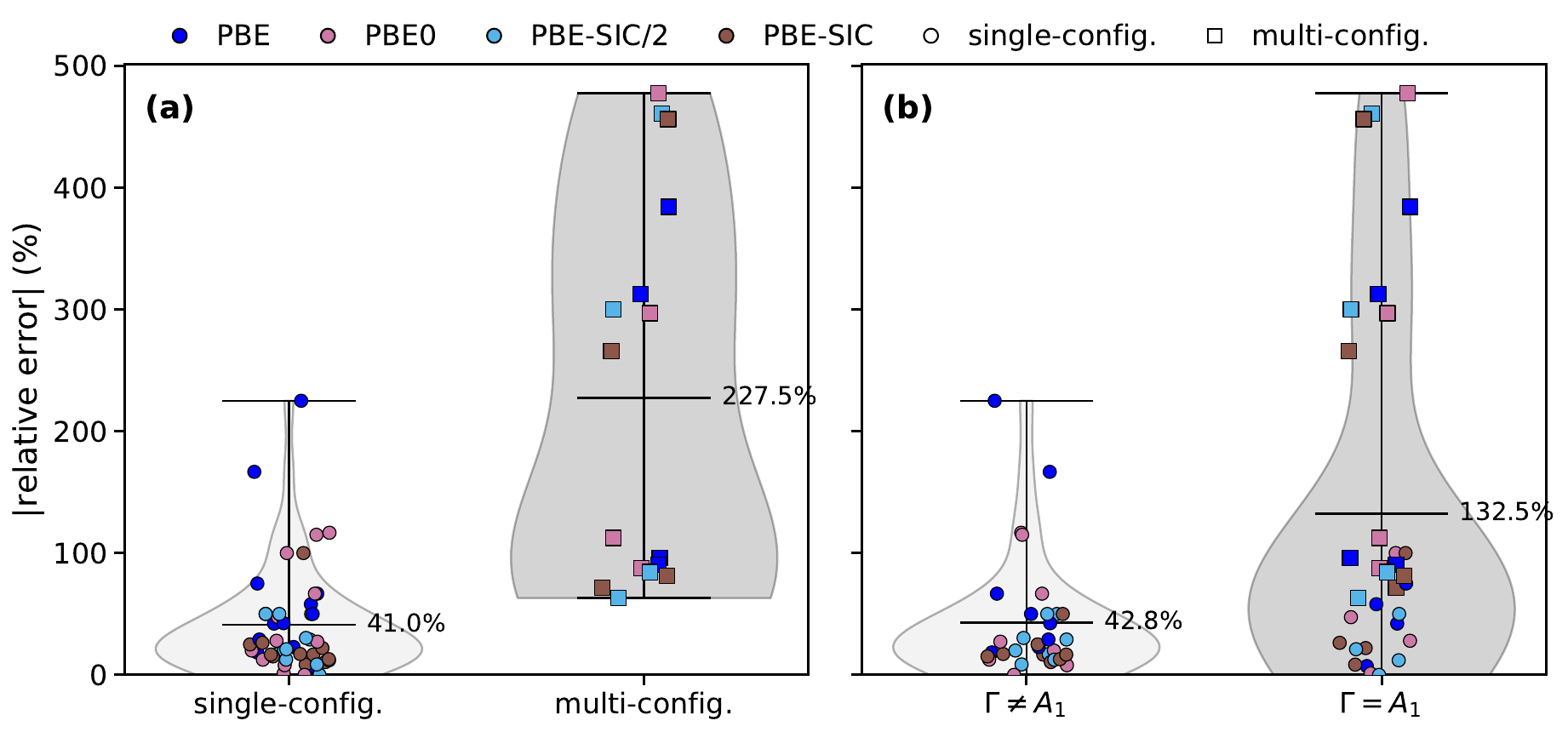}
    \caption{Relative errors in oscillator strength from orbital-optimized calculations, grouped according to (a) the single- or multi-configurational character of the excited states and (b) their irreducible representation relative to that of the ground state. States belonging to the totally symmetric representation ($A_1$) have nonzero overlap with the ground state. The mean value is indicated by a horizontal bar. Only states with $f_{\mathrm{ref}} > 0.001$ are included. The largest errors are associated primarily with multi-configurational states rather than with nonzero overlap between the ground and excited state.}
    \label{fig:overlap_error}
\end{figure}
To confirm that the large errors in the OO calculations are associated with the multi-configurational states and not with large overlaps with the ground state, Figure~\ref{fig:overlap_error} shows the errors in oscillator strength sorted into subgroups of transitions with single- and multi-configurational character. It also distinguishes excited states belonging to the totally symmetric irreducible representation, $\Gamma=A_1$, which have a nonzero overlap with the ground state, from those with $\Gamma \neq A_1$, which are orthogonal to the ground state by symmetry.
On one hand, for the single-configurational states, the errors are narrowly distributed, with a mean absolute value of $\sim$40\%. On the other hand, all multi-configurational states have errors above 50\%, with a mean absolute error above 200\%. The correlation of the error with the nonzero overlap between the ground and the excited state is less clear. States belonging to the totally symmetric representation have a larger mean error and a wider spread than those with different symmetry than the ground state. However, among the states with nonzero overlap, those with the largest deviation with the reference value of oscillator strength are the states with multi-configurational character, which turn out also to have $A_1$ symmetry. Single-configurational states with $A_1$ symmetry are generally affected by a smaller error.

\begin{figure}[hbt]
    \centering
    \includegraphics[width=\linewidth]{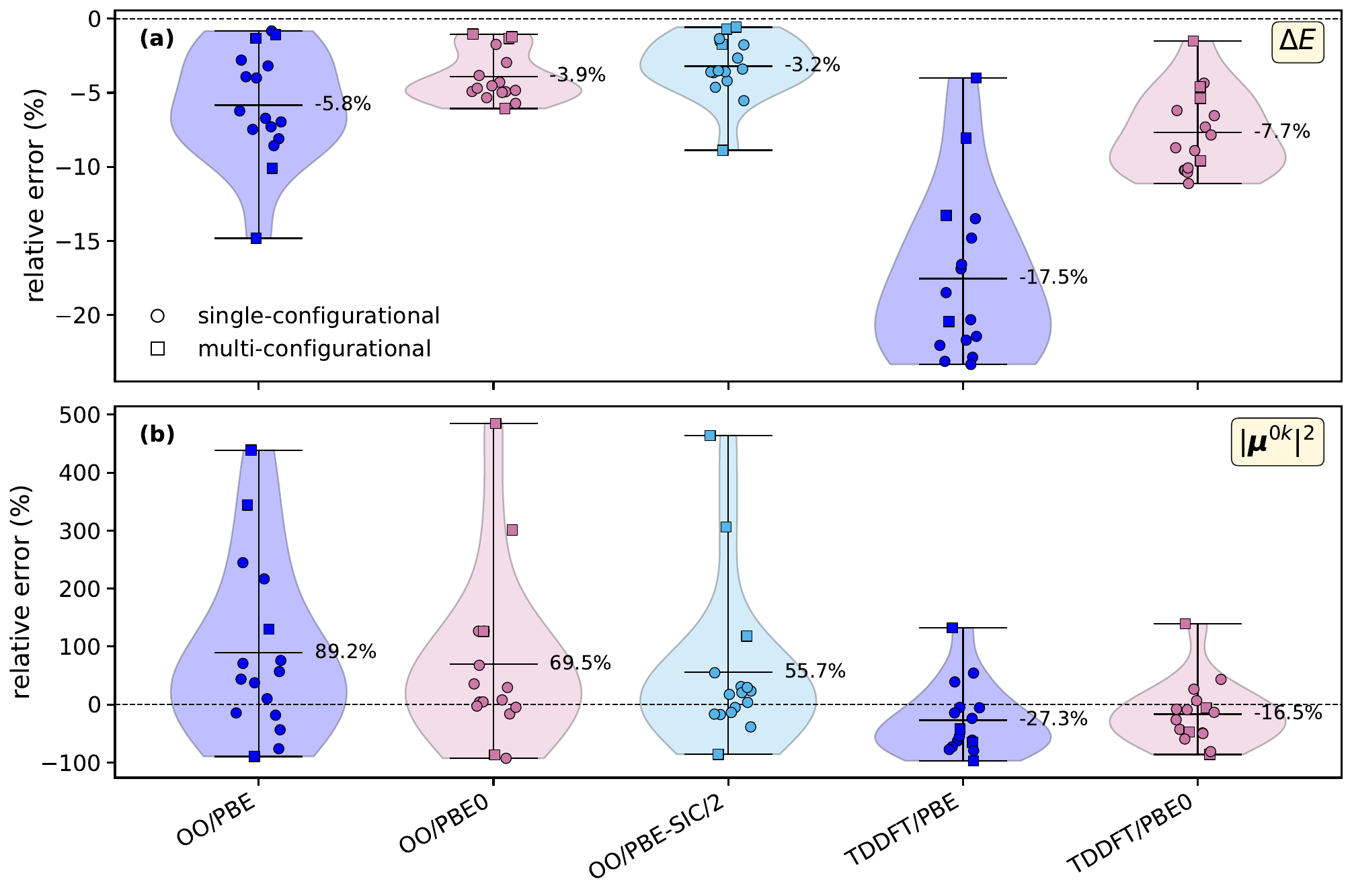}
    \caption{Signed relative errors on (a) the excitation energy and (b) the squared magnitude of the transition dipole moment for orbital-optimized and linear-response time-dependent density functional theory calculations using various exchange-correlation functionals. The mean value is indicated by a horizontal bar. Only states with $f_{\mathrm{ref}} > 0.001$ are included. Excited states with multi-configurational character are marked with squares, those with single-configurational character with circles. The OO calculations are affected by smaller errors on the excitation energy, but larger errors on the transition dipole moment, especially for the multi-configurational states.}
    \label{fig:error_energy_vs_TDM}
\end{figure}

Figure~\ref{fig:error_energy_vs_TDM} shows the distributions of errors on the excitation energy, $\Delta E$, and on the squared magnitude of the \gls*{tdm}, $\vert \bm{\mu} ^{0k}_{\mathrm{S}}\vert ^2$, separately, which both enter the expression for the calculation of the oscillator strength, according to eq\ \eqref{eq:osc_strength}.
The excitation energy is systematically underestimated by all methods. The \gls*{oo} calculations yield significantly smaller mean deviation and narrower distributions compared to \gls*{lrdft}. This difference is particularly large for PBE, for which the signed mean error is approximately $-6\%$ with \gls*{oo}, compared with $\sim-18\%$ with \gls*{lrdft}. The opposite trend is found for the squared magnitude of the \gls*{tdm}. The \gls*{oo} results exhibit larger mean errors and wider distributions than \gls*{lrdft}, with predominantly positive deviations, which suggests a systematic overestimation of $\vert \bm{\mu} ^{0k}_{\mathrm{S}}\vert ^2$. The OO distributions feature outliers with large errors corresponding to the multi-configurational states. In comparison, the \gls*{lrdft} results display narrower spread and smaller mean signed errors, with a negative bias that points to a systematic underestimation of the transition dipole moment. The same analysis was repeated after removing states with multi-configurational character, and the results are shown in Figure~S6 of the \gls*{si}. The errors in $\Delta E$ remain essentially unchanged for the \gls*{oo} calculations. The largest change occurs for the squared magnitude of the \gls*{tdm}. In particular, the mean errors for \gls*{oo}/PBE and \gls*{oo}/PBE0 decrease from approximately 90\% and 70\% to 50\% and 24\%, respectively, approaching the LR-TDDFT mean errors of $-30\%$ and $-22\%$ for PBE and PBE0, respectively, more closely. Overall, this analysis shows that the multi-configurational character affects the OO-calculated TDMs much more strongly than the excitation energy. Consequently, the errors in the \gls*{oo} oscillator strengths arise predominantly from errors on the magnitude of the \gls*{tdm}, whereas errors on the excitation energy play a minor role. 

The analyses using absolute errors in absolute units in Figures~S7--S10 of the \gls*{si} confirm the trends and conclusions drawn from the relative percentage errors. In particular, the largest errors are associated with the multi-configurational states, whereas no correlation is found with the nonorthogonality between the ground and excited states.

\section{Discussion}
The present results show that oscillator strengths provide a stringent test of \gls*{oo} density functional calculations for excited states. As commonly done, here the transition dipole moment has been evaluated directly from the corresponding Kohn--Sham wave functions. While the energy is stationary with respect to wave function variations, the TDM is not. Wave function errors therefore contribute only at second order for the excitation energy but already at first order for the TDM. Consequently, errors on the oscillator strengths are expected to arise primarily from the TDM rather than from the excitation energy, which is consistent with the error analysis presented above, and the oscillator strength is expected to be more strongly affected by the choice of basis set and xc functional.

The results identify the configurational character as the main factor governing
the reliability of OO transition intensities. For predominantly
single-configurational states, good results are obtained, with the accuracy improving with the inclusion of exact exchange and self-interaction correction in the functional. The mean absolute relative error in oscillator strength decreases from approximately 67\% with PBE to 45\% with PBE0 and 25\% with PBE-SIC/2, the self-interaction-corrected OO calculations yielding smaller errors than LR-TDDFT/PBE0. These results support the
use of OO density functional calculations for both valence and Rydberg
transitions when a single optimized configuration provides an adequate
description. In contrast, large errors persist across all functionals
considered for states with substantial configurational mixing, showing
that improving the functional alone is insufficient for these states.
Nonorthogonality between independently optimized states is not found to
be an important source of error in this benchmark. These findings
identify the incorporation of configurational mixing, while retaining
state-specific orbital relaxation, as the principal methodological
development needed to address the largest failures.

Basis set effects are most significant for the most diffuse, higher-energy Rydberg states. These transitions generally have small oscillator strengths. Therefore, large relative differences can correspond to small absolute changes. Nevertheless, the increased basis set sensitivity of diffuse Rydberg states is also apparent from the absolute changes for several of the states considered here. In contrast, bright valence and Rydberg states with less diffuse character are much less sensitive to the basis set representation. Basis set effects thus do not appear to be critical for the oscillator strength. In contrast, as shown in the accompanying article~\cite{Restaino2026-iv}, the choice of basis set is more critical for the single-state dipole moment. Rydberg states can possess sizable dipole moments, and for the most diffuse states, even a doubly augmented atomic basis set may not be sufficient.

Large errors on the oscillator strength persist for a few states across all functionals, with no systematic improvement from PBE to PBE0 or PBE with \gls*{sic}. High-level wave function-based calculations from the literature identify these states as having large multi-configurational character. A clear example is provided by the S$_3$ and S$_4$ states of water. MS-CASPT2 calculations describe them as approximately 70/30 and 30/70 mixtures, respectively, of the 2p$_z \rightarrow$ 3s and 2p$_x \rightarrow$ 3p$_x$ excitations~\cite{Rubio2008-lk} and having similar oscillator strength. In the \gls*{oo} calculations, however, the two states are represented as single configurations, leading to incorrect transition intensities. The S$_3$ oscillator strength is strongly overestimated, whereas that of S$_4$ is almost vanishing. The absence of configurational mixing does not significantly affect the excitation energy of these states or, as shown in the accompanying article, their permanent dipole moment~\cite{Restaino2026-iv}. This agreement is fortuitous, as the two underlying single configurations happen to have similar excitation energy and dipole moment, so their mixing produces only small changes in these quantities~\cite{Levi2026-ts,Sigurdarson2023-do}. The oscillator strength is much more sensitive because it depends on the transition dipole moment, which is considerably different for the two configurations. Similar failures are found here for the S$_5$ and S$_1$ $\pi \rightarrow \pi^*$ states of formaldehyde and ethylene, respectively, both of which are known to exhibit strong configuration mixing in correlated wave function calculations~\cite{Gomez-Carrasco2010-me,Schreiber2008-ui,Muller2001-hq,Merchan1995-gc,Feller2014-ih,Sauri2011-au,Angeli2009-bv,Finley1998-on,Serrano-Andres1993-fn}. In particular, the S$_5$ state of formaldehyde lies close to two conical intersections at the Franck--Condon geometry~\cite{Gomez-Carrasco2010-me}, 
making this state especially challenging and explaining the large deviation in oscillator strength observed for this state. These cases highlight a fundamental limitation of state-of-the-art \gls*{oo} approaches. When each excited state is represented by a single determinant, configurational mixing does not arise naturally through orbital relaxation alone. Indeed, for the S$_3$ and S$_4$ states of water, scanning the energy as a function of the two orbital rotations that mix the 3s and 3p$_x$ orbitals did not reveal any solution with mixed configurational character. This indicates that improving the exchange--correlation functional alone cannot resolve these failures when the dominant error originates from the
lack of static correlation.

Once the multi-configurational states are excluded, a clear functional dependence emerges. PBE0 and PBE with \gls*{sic} improve the TDM and oscillator strength significantly over PBE, with PBE-SIC/2 yielding the smallest mean absolute relative errors ($\sim$25\% for the oscillator strength). \Gls*{sic}, which restores the correct $-1/r$ asymptotic dependence of the effective potential, generally improves the TDM in the present calculations. Similar improvements have been found previously for the excitation energy~\cite{Sigurdarson2023-do}. However, in the accompanying article, it is found that \gls*{sic} has the opposite effect on excited-state dipole moments, as it tends to overestimate their magnitude~\cite{Restaino2026-iv}. This is attributed to a tendency of globally scaled \gls*{sic} to overcorrect in regions where occupied orbital densities overlap. Locally scaled \gls*{sic}, which is currently under development~\cite{John2026arXiv}, may provide a more balanced treatment of the excitation energy, dipole moment and transition intensities.

The \gls*{lrdft} spectra are generally significantly red-shifted relative to the \gls*{oo} and coupled-cluster spectra, reflecting the well-known tendency of \gls*{lrdft} to underestimate the excitation energy, particularly for diffuse Rydberg states. In contrast, \gls*{lrdft} with PBE and PBE0 is found generally to yield smaller errors on the TDM than \gls*{oo} calculations with the same functionals. The improvement is most significant for the states with multi-configurational character. The better performance of \gls*{lrdft} for transition intensities might be the result of the fact that in \gls*{lrdft}, an excited state is represented as a linear combination of single excitations, allowing for a mixing of configurations that is absent in the \gls*{oo} calculations, which are based on a single optimized determinant. However, the improved oscillator strengths do not always reflect a correct description of the excited states. For water, for example, the high-energy peak in the spectrum arises almost entirely from the S$_3$ transition in the \gls*{lrdft} calculations, whereas in the MS-CASPT2 calculations both the S$_3$ and S$_4$ states contribute with similar intensity, with each state exhibiting strong mixing of the $2p_z \rightarrow 3s$ and $2p_x \rightarrow 3p_x$ configurations.
In formaldehyde, several states with $\pi \rightarrow \pi^*$ character are obtained in \gls*{lrdft} calculations, and the intensity of this transition is underestimated compared to higher-level results if the state with the largest contribution of $\pi \rightarrow \pi^*$ excitation is considered. Thus, it is found that although \gls*{lrdft} generally improves the TDM and oscillator strength for the excited states of small molecules considered here, particularly for multi-configurational states, it does not always provide an accurate description of their electronic character, as it may give an unphysical mixing of excitations. Qualitatively incorrect state mixing has previously been reported in adiabatic
\gls*{lrdft} calculations for other systems~\cite{Selenius2024-rk,Taka2022,Toffoli2022},
including charge-transfer excitations~\cite{Selenius2024-rk}, and has been
attributed to insufficient orbital relaxation. 

It has previously been suggested that a nonzero overlap between the ground and excited states with the same symmetry could artificially inflate the transition dipole moment in OO calculations~\cite{Gilbert2008, Bourne-Worster2021-hb}. This potential source of error on the computed oscillator strengths is also investigated here. Tables~S5 to S8 in the \gls*{si} list the overlaps between the calculated \gls*{oo} excited states and the ground state. These results do not show a clear correlation between the magnitude of the overlap and the error on the oscillator strength, as also found in Figure\ \ref{fig:overlap_error}. For example, for water, the overlap between the ground state and the problematic S$_3$ state vanishes when using PBE0 and is greatly reduced with PBE-SIC/2 compared to PBE. However, the oscillator strength is largely overestimated with all functionals, as the main source of error for this state is the lack of a multi-configurational character. For ammonia, the overlap between the ground and S$_1$ states, which has a dominant single-configurational character, is nonzero for all xc functionals and becomes larger in calculations that employ \gls*{sic}. Yet, the oscillator strength stays close to the coupled-cluster value for all functionals.

\section{Conclusions}
In this work, orbital-optimized density functional calculations are carried out to compute oscillator strengths and absorption spectra including valence and Rydberg excited states up to 10 eV for water, formaldehyde, ammonia, methanol, and ethylene. The calculations employ a plane-wave basis set to describe diffuse Rydberg orbitals accurately, and the computation of the transition dipole moment is performed by using the Kohn-Sham wave function with nonorthogonal matrix elements formulated within the projector augmented wave approach. 

All functionals yield relatively small errors on the excitation energy and hence the position of the peaks in the spectrum, compared to results from higher-level coupled-cluster calculations with atomic orbitals basis sets including sufficiently diffuse functions. The relatively simple GGA functional PBE gives a mean relative error of $\sim-6\%$, whereas PBE0 and self-interaction-corrected PBE reduce it to between $-4\%$ and $-3\%$. In contrast, linear-response time-dependent density functional theory calculations produce significantly larger red shifts, especially for transitions to diffuse Rydberg excited states, with a mean error approximately three and two times larger than for OO calculations with PBE and PBE0, respectively.

For some excited states, the OO calculations exhibit systematically large errors in the oscillator strength that are found to be largely independent of the exchange-correlation functional. The source of these large errors is traced back to the fact that these states should have a large multi-configurational character based on high-level correlated wave function calculations, but that is not captured by the OO calculations, as they rely on the optimization of a single determinant. As a result, the transition dipole moment is poorly described for these excitations. In contrast, nonzero overlap between the ground and excited states with the same symmetry is not found to be a significant source of error. For states with single-configurational character, the OO calculations generally provide reliable transition dipole moments and oscillator strengths, although the errors remain slightly larger than those obtained with \gls*{lrdft} using the same functional. PBE0 and self-interaction-corrected PBE, which improve the long-range behavior of the potential, perform better than PBE, with PBE-SIC/2 yielding the smallest mean absolute relative error on the oscillator strength, approximately 25\%. Compared with OO calculations, \gls*{lrdft} gives smaller errors on the transition dipole moment and oscillator strength for multi-configurational states by allowing the mixing of different single excitations. In some cases, however, this mixing differs from that predicted by higher-level calculations, indicating that the improved transition intensity does not necessarily reflect an accurate description of the electronic character of the excited state. 

Several methodological developments could further improve the calculation of spectra within the \gls*{oo} density functional approach. The restricted open-shell Kohn--Sham (ROKS) method~\cite{Filatov1998, Frank1998} allows singlet excited states to be targeted directly, avoiding the separate mixed-spin and triplet calculations required by the post-SCF spin-purification approach employed here with spin-unrestricted calculations. ROKS would also facilitate the inclusion of vibronic effects in simulated spectra, since analytical excited-state gradient and Hessian can be evaluated more readily. Although ROKS has been available for several decades, its performance in the prediction of spectra beyond low-lying valence excitations remains largely unexplored. Most importantly, OO calculations will need to incorporate multi-configurational character while retaining state-specific orbital relaxation in order to provide a reliable description of transition properties for strongly mixed excited states. Recent approaches based on fractional occupation of the orbitals offer a promising route~\cite{Sinyavskiy2025}. In this formulation, the occupations are derived from \gls*{lrdft} excitation amplitudes. The dependence on \gls*{lrdft} may be problematic however, because adiabatic \gls*{lrdft} can produce qualitatively incorrect mixing between excitations, as shown here and in previous works~\cite{Selenius2024-rk, Taka2022}. Developing an extension of the OO density functional method that can represent multi-configurational excited states while retaining the state-specific orbital optimization is an important direction for future work, which we are currently pursuing.

\appendix

\section{Derivation of the transition dipole moment in the projector augmented wave formalism}
\label{sec:paw_nonorthogonal_matrix_elements}

\subsection{One-electron matrix elements}
In the PAW approach\cite{Blochl1994-gk}, the matrix elements of a one-electron operator $\hat O$ between sets of independently optimized orbitals belonging to states $k$ and $k^\prime$ are obtained using the linear transformation defined in eq\ \eqref{eq:paw_transform}~\cite{Blochl1994-gk}. When using the frozen-core approximation for the orbitals, as commonly done\cite{Mortensen2024-ji, Blochl2002}, valence--valence, valence--core, and core--core matrix elements need to be considered:
\begin{align}
O_{ij}^{kk^\prime}
&=
\expv{\psi_i^k}{\hat O}{\psi_j^{k^\prime}}
=
\expv{\tilde{\psi}_i^k}
{\hat{\mathcal T}^{\dagger}\hat O\hat{\mathcal T}}
{\tilde{\psi}_j^{k^\prime}},
\label{eq:paw_vv_matrix_element}
\\
O_{ic}^{k}
&=
\expv{\psi_i^k}{\hat O}{\phi_c^{\mathrm{core}}}
= \expv{\tilde{\psi}_i^k}{\hat{\mathcal T}^{\dagger}\hat O}{\phi_c^{\mathrm{core}}},
\label{eq:paw_vc_matrix_element}
\\
O_{cj}^{k^\prime}
&=
\expv{\phi_c^{\mathrm{core}}}{\hat O}{\psi_j^{k^\prime}}
= \expv{\phi_c^{\mathrm{core}}}{\hat O\hat{\mathcal T}}
{\tilde{\psi}_j^{k^\prime}},
\label{eq:paw_cv_matrix_element}
\\
O_{cd}
&=
\expv{\phi_c^{\mathrm{core}}}{\hat O}
{\phi_d^{\mathrm{core}}}.
\label{eq:paw_cc_matrix_element}
\end{align}

First, the following one-center expansions of the orbitals and pseudo orbitals inside the augmentation sphere of atom $a$ are defined, 
\begin{equation}
\ket{\psi_i^{a,k}}
=
\sum_n
\ket{\phi_n^a}
P_{in}^{a,k},
\qquad
\ket{\tilde{\psi}_i^{a,k}}
=
\sum_n
\ket{\tilde{\phi}_n^a}
P_{in}^{a,k},
\end{equation}
where $P_{in}^{a,k}=\bk{\tilde p_n^a}{\tilde{\psi}_i^k}$ are projector overlaps. Each orbital can then be written as
\begin{equation}
\ket{\psi_i^k}
=
\ket{\tilde{\psi}_i^k}
+
\sum_a
\left(
\ket{\psi_i^{a,k}}
-
\ket{\tilde{\psi}_i^{a,k}}
\right).
\end{equation}
Substituting the atomic expansion in eq\ \eqref{eq:paw_vv_matrix_element} for the valence-valence matrix elements gives
\begin{align}
O_{ij}^{kk^\prime}
={}&
\expv{\tilde{\psi}_i^k}{\hat O}{\tilde{\psi}_j^{k^\prime}}
\nonumber\\
&+
\sum_a
\expv{\psi_i^{a,k}-\tilde{\psi}_i^{a,k}}
{\hat O}
{\tilde{\psi}_j^{k^\prime}}
\nonumber\\
&+
\sum_{a^\prime}
\expv{\tilde{\psi}_i^k}
{\hat O}
{\psi_j^{a^\prime,k^\prime}
-\tilde{\psi}_j^{a^\prime,k^\prime}}
\nonumber\\
&+
\sum_{aa^\prime}
\expv{\psi_i^{a,k}-\tilde{\psi}_i^{a,k}}
{\hat O}
{\psi_j^{a^\prime,k^\prime}
-\tilde{\psi}_j^{a^\prime,k^\prime}}.
\label{eq:paw_valence_expansion}
\end{align}
Writing each pseudo orbital as
$\ket{\tilde{\psi}_i^k}
=
\ket{\tilde{\psi}_i^{a,k}}
+
\left(
\ket{\tilde{\psi}_i^k}
-
\ket{\tilde{\psi}_i^{a,k}}
\right)$,
eq~\ref{eq:paw_valence_expansion} can be rearranged as
\begin{align}
O_{ij}^{kk^\prime}
={}&
\expv{\tilde{\psi}_i^k}{\hat O}{\tilde{\psi}_j^{k^\prime}}
\nonumber\\
&+
\sum_a
\left[
\expv{\psi_i^{a,k}}{\hat O}{\psi_j^{a,k^\prime}}
-
\expv{\tilde{\psi}_i^{a,k}}{\hat O}
{\tilde{\psi}_j^{a,k^\prime}}
\right]
\nonumber\\
&+
\sum_a
\expv{\psi_i^{a,k}-\tilde{\psi}_i^{a,k}}
{\hat O}
{\tilde{\psi}_j^{k^\prime}
-\tilde{\psi}_j^{a,k^\prime}}
\nonumber\\
&+
\sum_a
\expv{\tilde{\psi}_i^k-\tilde{\psi}_i^{a,k}}
{\hat O}
{\psi_j^{a,k^\prime}
-\tilde{\psi}_j^{a,k^\prime}}
\nonumber\\
&+
\sum_{a\neq a^\prime}
\expv{\psi_i^{a,k}-\tilde{\psi}_i^{a,k}}
{\hat O}
{\psi_j^{a^\prime,k^\prime}
-\tilde{\psi}_j^{a^\prime,k^\prime}}.
\label{eq:paw_valence_decomposition}
\end{align}
For a local operator, the last three terms in eq\ \eqref{eq:paw_valence_decomposition} vanish when the expansions in partial waves are converged. The differences $\psi_i^{a,k}-\tilde{\psi}_i^{a,k}$ are confined to the corresponding augmentation spheres, while $\tilde{\psi}_i^k-\tilde{\psi}_i^{a,k}$ vanishes within the same regions. In addition, augmentation spheres centered on different atoms do not overlap. The valence--valence matrix elements therefore reduce to
\begin{align}
O_{ij}^{kk^\prime}
={}&
\expv{\tilde{\psi}_i^k}{\hat O}{\tilde{\psi}_j^{k^\prime}}
+
\sum_a
\left[
\expv{\psi_i^{a,k}}{\hat O}{\psi_j^{a,k^\prime}}
-
\expv{\tilde{\psi}_i^{a,k}}{\hat O}
{\tilde{\psi}_j^{a,k^\prime}}
\right].
\end{align}
Expanding the one-center terms in partial waves gives
\begin{align}
O_{ij}^{kk^\prime}
={}&
\expv{\tilde{\psi}_i^k}{\hat O}{\tilde{\psi}_j^{k^\prime}}
+
\sum_a\sum_{nm}
\left[
\expv{\phi_n^a}{\hat O}{\phi_m^a}
-
\expv{\tilde{\phi}_n^a}{\hat O}{\tilde{\phi}_m^a}
\right]
D_{nm,ij}^{a,kk^\prime},
\label{eq:paw_vv_final}
\end{align}
where
\begin{equation}
D_{nm,ij}^{a,kk^\prime}
=
P_{in}^{a,k*}P_{jm}^{a,k^\prime}
=
\bk{\tilde{\psi}_i^k}{\tilde p_n^a}
\bk{\tilde p_m^a}{\tilde{\psi}_j^{k^\prime}}.
\end{equation}

For a frozen-core orbital $\phi_c^{\mathrm{core}}$ centered on atom $a$, the valence--core matrix element is
\begin{align}
O_{ic}^{k}
={}&
\expv{\tilde{\psi}_i^k}
{\hat O}
{\phi_c^{\mathrm{core}}}
+
\sum_a
\expv{\psi_i^{a,k}-\tilde{\psi}_i^{a,k}}
{\hat O}
{\phi_c^{\mathrm{core}}}.
\label{eq:paw_vc_expansion}
\end{align}
Since each frozen-core orbital is localized inside the augmentation sphere, and within augmentation spheres
$\tilde{\psi}_i^k=\tilde{\psi}_i^{a,k}$, eq\ \eqref{eq:paw_vc_expansion} becomes
\begin{align}
O_{ic}^{k}
&=
\expv{\tilde{\psi}_i^{a,k}}
{\hat O}
{\phi_c^{\mathrm{core}}}
+
\expv{\psi_i^{a,k}-\tilde{\psi}_i^{a,k}}
{\hat O}
{\phi_c^{\mathrm{core}}}
\nonumber\\
&=
\expv{\psi_i^{a,k}}
{\hat O}
{\phi_c^{\mathrm{core}}}
\nonumber\\
&=
\sum_n
P_{in}^{a,k*}
\expv{\phi_n^{a}}
{\hat O}
{\phi_c^{\mathrm{core}}},
\label{eq:paw_valence_core_matrix_element}
\end{align}
where the core orbital $c$ is localized in the augmentation sphere of atom $a$. Analogously,
\begin{equation}
O_{cj}^{k^\prime}
=
\sum_m
\expv{\phi_c^{\mathrm{core}}}
{\hat O}
{\phi_m^{a}}
P_{jm}^{a,k^\prime}.
\label{eq:paw_core_valence_matrix_element}
\end{equation}

The valence--core and core--valence matrix elements, as well as the core--core matrix elements, do not necessarily vanish. Their contributions to the many-electron matrix element $\langle \Psi ^k | \hat O | \Psi ^{k^{\prime}}\rangle$ and to the TDM are instead determined by the corresponding cofactors of the orbital overlap matrix, as shown below.

\subsection{Transition dipole moment}
To derive the final expression of the TDM between nonorthogonal OO states in the PAW formalism, we also need to consider the cofactor of the overlap matrix between orbitals (see eqs\ \eqref{Lowdin_NOCI} and \eqref{eq:Lowdin_NOCI_unrestr_final}).

The frozen-core orbitals are identical for states $k$ and $k^\prime$ and are orthonormal among themselves. Moreover, the core orbitals and the partial waves are commonly obtained from the same atomic KS calculation~\cite{Blochl2002}, as in the GPAW program~\cite{Mortensen2024-ji}. The partial waves are therefore usually orthogonal to the core orbitals, for each atom $a$. Thus, the overlap between a core and a valence orbital vanishes (see eq\ \eqref{eq:paw_valence_core_matrix_element}),
\begin{align}
S_{ic}^{k}
=
\bk{\psi_i^k}{\phi_c^{\mathrm{core}}}
&=
\sum_n
P_{in}^{a,k*}
\bk{\phi_n^a}{\phi_c^{\mathrm{core}}}
=
0.
\end{align}
Together with the orthonormality of the frozen-core orbitals, this gives
\begin{equation}
S_{ic}^{k}
=
0,
\qquad
S_{cj}^{k^\prime}
=
0,
\qquad
S_{cd}
=
\bk{\phi_c^{\mathrm{core}}}{\phi_d^{\mathrm{core}}}
=
\delta_{cd}.
\end{equation}
Therefore, for each spin component $\sigma$, the complete overlap matrix between occupied orbitals of the two OO states, $k$ and $k^\prime$, has the block-diagonal form
\begin{equation}
\mathbf S_{\sigma,\mathrm{full}}^{kk^\prime}
=
\begin{pmatrix}
\mathbf S_{\sigma}^{kk^\prime} & 0\\
0 & \mathbf I_{\sigma,\mathrm{core}}
\end{pmatrix},
\label{eq:paw_full_overlap}
\end{equation}
where $\mathbf S_{\sigma}^{kk^\prime}$ contains only the overlaps between valence orbitals. Its determinant and inverse are
\begin{equation}
\det\!\left(
\mathbf S_{\sigma,\mathrm{full}}^{kk^\prime}
\right)
=
\det\!\left(
\mathbf S_{\sigma}^{kk^\prime}
\right),
\qquad
\left(
\mathbf S_{\sigma,\mathrm{full}}^{kk^\prime}
\right)^{-1}
=
\begin{pmatrix}
\left(
\mathbf S_{\sigma}^{kk^\prime}
\right)^{-1} & 0\\
0 & \mathbf I_{\sigma,\mathrm{core}}
\end{pmatrix}.
\end{equation}
Using $ \operatorname{cof}\!\left(\mathbf S\right)_{ij} = \operatorname{adj}\!\left(\mathbf S\right)_{ji} = \det\!\left(\mathbf S\right) \left(\mathbf S^{-1}\right)_{ji}$, the cofactor matrix of the complete overlap matrix is
\begin{equation}
\operatorname{cof}\!\left(
\mathbf S_{\sigma,\mathrm{full}}^{kk^\prime}
\right)
=
\begin{pmatrix}
\operatorname{cof}\!\left(
\mathbf S_{\sigma}^{kk^\prime}
\right)
&
0
\\[2mm]
0
&
\det\!\left(
\mathbf S_{\sigma}^{kk^\prime}
\right)
\mathbf I_{\sigma,\mathrm{core}}
\end{pmatrix}.
\label{eq:paw_full_cofactor}
\end{equation}
Eq\ \eqref{eq:paw_full_cofactor} shows that the cofactors multiplying the valence--core and core--valence matrix elements vanish. Within the core--core block, only the diagonal elements have nonzero cofactors. Thus, although the corresponding one-electron integrals do not necessarily vanish, the mixed valence--core and off-diagonal core--core contributions do not enter the many-electron matrix element.

Applying eq\ \eqref{eq:paw_full_cofactor} together with the expressions of the one-electron matrix elements obtained in the previous section in L{\"o}wdin's expression with the dipole moment operator, eq\ \eqref{eq:Lowdin_NOCI_unrestr_final}, the transition dipole moment between spin-unrestricted nonorthogonal determinants becomes
\begin{align}
\expv{\Psi^k}{\hat{\bm{\mu}}}{\Psi^{k^\prime}}
={}&
-\det\!\left(\mathbf S_{\beta}^{kk^\prime}\right)
\sum_{ij\in\alpha}
\mathbf r_{ij}^{kk^\prime}
\operatorname{cof}\!\left(
\mathbf S_{\alpha}^{kk^\prime}
\right)_{ij}
\nonumber\\
&-
\det\!\left(\mathbf S_{\alpha}^{kk^\prime}\right)
\sum_{ij\in\beta}
\mathbf r_{ij}^{kk^\prime}
\operatorname{cof}\!\left(
\mathbf S_{\beta}^{kk^\prime}
\right)_{ij}
\nonumber\\
&+
\det\!\left(\mathbf S_{\alpha}^{kk^\prime}\right)
\det\!\left(\mathbf S_{\beta}^{kk^\prime}\right)
\left[
\sum_a
\mathcal Z_a\mathbf R_a
-
\sum_c
\expv{\phi_c^{\mathrm{core}}}
{\mathbf r}
{\phi_c^{\mathrm{core}}}
\right],
\label{eq:paw_tdm_with_core}
\end{align}
where $c$ runs over all occupied frozen-core spin orbitals. The mixed valence--core and core--valence matrix elements as well as the off-diagonal core--core matrix elements are absent because their corresponding cofactors vanish. The surviving valence--valence position matrix elements are obtained from eq\ \eqref{eq:paw_vv_final} as
\begin{align}
\mathbf r_{ij}^{kk^\prime}
=
\expv{\psi_i^k}{\mathbf r}{\psi_j^{k^\prime}}
={}&
\expv{\tilde{\psi}_i^k}{\mathbf r}{\tilde{\psi}_j^{k^\prime}}
+
\sum_a\sum_{nm}
\left[
\expv{\phi_n^a}{\mathbf r}{\phi_m^a}
-
\expv{\tilde{\phi}_n^a}{\mathbf r}{\tilde{\phi}_m^a}
\right]
D_{nm,ij}^{a,kk^\prime}.
\label{eq:paw_position_appendix}
\end{align}
Since typically the frozen-core densities are symmetric around each atomic center by construction,
\begin{equation}
\sum_c
\expv{\phi_c^{\mathrm{core}}}
{\mathbf r}
{\phi_c^{\mathrm{core}}}
=
\sum_a
N_{\mathrm{core}}^a\mathbf R_a,
\label{eq:paw_core_position}
\end{equation}
where $N_{\mathrm{core}}^a$ is the number of frozen-core electrons associated with atom $a$. Substituting eq\ \eqref{eq:paw_core_position} into eq\ \eqref{eq:paw_tdm_with_core} and combining the electronic core contribution with the nuclear term, yields
\begin{align}
\expv{\Psi^k}{\hat{\bm{\mu}}}{\Psi^{k^\prime}}
={}&
-\det\!\left(\mathbf S_{\beta}^{kk^\prime}\right)
\sum_{ij\in\alpha}
\mathbf r_{ij}^{kk^\prime}
\operatorname{cof}\!\left(
\mathbf S_{\alpha}^{kk^\prime}
\right)_{ij}
\nonumber\\
&-
\det\!\left(\mathbf S_{\alpha}^{kk^\prime}\right)
\sum_{ij\in\beta}
\mathbf r_{ij}^{kk^\prime}
\operatorname{cof}\!\left(
\mathbf S_{\beta}^{kk^\prime}
\right)_{ij}
\nonumber\\
&+
\det\!\left(\mathbf S_{\alpha}^{kk^\prime}\right)
\det\!\left(\mathbf S_{\beta}^{kk^\prime}\right)
\sum_a
\left(
\mathcal Z_a-N_{\mathrm{core}}^a
\right)
\mathbf R_a,
\end{align}
which is the expression of the TDM between nonorthogonal OO states in the PAW formalism reported in eq\ \eqref{eq:Lowdin_NOCI_unrestr_final_paw}.

The overlaps between valence orbitals entering $\mathbf S_{\alpha}^{kk^\prime}$ and $\mathbf S_{\beta}^{kk^\prime}$ are obtained by setting $\hat O=1$ in eq\ \eqref{eq:paw_vv_final},
\begin{align}
S_{ij}^{kk^\prime}
={}&
\bk{\tilde{\psi}_i^k}{\tilde{\psi}_j^{k^\prime}}
+
\sum_a\sum_{nm}
\left[
\bk{\phi_n^a}{\phi_m^a}
-
\bk{\tilde{\phi}_n^a}{\tilde{\phi}_m^a}
\right]
D_{nm,ij}^{a,kk^\prime}.
\label{eq:paw_overlap_matrix_element}
\end{align}
For \gls*{pw} and real-space grid representations, this expression is evaluated directly from the smooth pseudo orbitals and the atom-centered PAW corrections. For an \gls*{lcao} representation, the overlap matrix for spin component $\sigma$ can equivalently be evaluated as
\begin{equation}
\mathbf S_{\sigma}^{kk^\prime}
=
\mathbf C_{\sigma}^{k\dagger}
\mathbf S^{\mathrm{AO}}
\mathbf C_{\sigma}^{k^\prime},
\label{eq:paw_overlap_lcao}
\end{equation}
where $\mathbf C_{\sigma}^k$ and $\mathbf C_{\sigma}^{k^\prime}$ are the matrices of coefficients of the pseudo orbitals for states $k$ and $k^\prime$, respectively, and $\mathbf S^{\mathrm{AO}}$ is the overlap matrix in the atomic orbital basis including the PAW corrections, 
\begin{align}
S_{\mu\nu}^{\mathrm{AO}}
={}&
\bk{\Phi_\mu}{\Phi_\nu}
+
\sum_a\sum_{nm}
\bk{\Phi_\mu}{\tilde p_n^a}
\left[
\bk{\phi_n^a}{\phi_m^a}
-
\bk{\tilde{\phi}_n^a}{\tilde{\phi}_m^a}
\right]
\bk{\tilde p_m^a}{\Phi_\nu},
\label{eq:paw_ao_overlap}
\end{align}
with $\Phi_\mu$ and $\Phi_\nu$ being the atomic orbital basis functions.

\section{Spin purification of the transition dipole moment}
\label{app:spin_purification_tdm}

In the approximation that the mixed-spin determinant and the corresponding spin-adapted singlet and triplet wave functions are constructed from the same set of spatial orbitals, the former is an equal mixture of the singlet and the $M_S=0$ component of the triplet. For an excited state $k$ (the doubly occupied orbitals, which are common to all wave functions, are omitted for simplicity),
\begin{equation}
\ket{\psi_i^k\bar{\psi}_j^k}
=
\frac{1}{\sqrt{2}}
\left(
\ket{{}^1\Psi^k}
+
\ket{{}^3\Psi^k}
\right),
\label{eq:mixed_spin_decomposition}
\end{equation}
where $\psi_i^k$ and $\bar{\psi}_j^k$ denote the singly occupied $\alpha$ and $\beta$ orbitals, respectively.
Under this approximation, the \gls*{tdm} between the mixed-spin excited state and the singlet ground state $\ket{{}^1\Psi^0}$ is
\begin{align}
\expv{\psi_i^k\bar{\psi}_j^k}
{\hat{\bm{\mu}}}
{{}^1\Psi^0}
={}&
\frac{1}{\sqrt{2}}
\left(
\expv{{}^1\Psi^k}
{\hat{\bm{\mu}}}
{{}^1\Psi^0}
+
\expv{{}^3\Psi^k}
{\hat{\bm{\mu}}}
{{}^1\Psi^0}
\right).
\label{eq:mixed_spin_tdm}
\end{align}
Since the electric dipole operator is spin-independent, the singlet--triplet matrix element vanishes in the absence of spin--orbit coupling, $\expv{{}^3\Psi^k} {\hat{\bm{\mu}}} {{}^1\Psi^0}=0$, and eq\ \eqref{eq:mixed_spin_tdm} gives the following expression for the TDM of the singlet excited state:
\begin{equation}
\expv{{}^1\Psi^k}
{\hat{\bm{\mu}}}
{{}^1\Psi^0}
=
\sqrt{2}\,
\expv{\psi_i^k\bar{\psi}_j^k}
{\hat{\bm{\mu}}}
{{}^1\Psi^0}.
\label{eq:spin_purified_tdm}
\end{equation}
Thus, the transition dipole moment of the pure singlet transition is larger by a factor of $\sqrt{2}$ than that obtained using the corresponding mixed-spin determinant. 

\begin{acknowledgement}
The authors thank Denis Jacquemin and Pierre-François Loos for providing the results of the CCSDT/\mbox{d-aug-cc-pVTZ} calculations of ammonia. The authors thank Aleksei V. Ivanov, Hannes Jónsson, and Philipp Hansmann for useful discussions. Y.L.A.S acknowledges support by the Max Planck Society. G.L. and D.L.P. acknowledge support by the Icelandic Research Fund (grant no. 2511544). E. Ö. J. acknowledges support by the Icelandic Research Fund (grant no. 2611846). G.L. acknowledges support from the ERC under the European Union's Horizon Europe research and innovation programme (grant no. 101166044, project NEXUS). Views and opinions expressed are however those of the author(s) only and do not necessarily reflect those of the European Union or ERC Executive Agency. Neither the European Union nor the granting authority can be held responsible for them. The authors acknowledge computer resources, data storage, and user support by the Icelandic Research e-Infrastructure (IREI), funded by the Icelandic Infrastructure Fund.
\end{acknowledgement}

\begin{suppinfo}
The Supporting Information includes numerical tests on the size of the simulation box; additional LR-TDDFT calculations with different LCAO basis sets; graphical depictions of the molecular orbitals involved in the excitations examined in the article; tables listing the overlap between the ground state and the orbital-optimized excited state for each molecule; and additional plots illustrating the analysis of the methodological error in terms of absolute errors.
\end{suppinfo}

\section*{Data and Software Availability}
The data underlying this study are available in the published article and the Supporting Information. Scripts and code used to generate the results are publicly available at \url{https://gitlab.com/lore_94/OO-excited-state-properties}.

\begin{tocentry}
   \includegraphics[width=\textwidth]{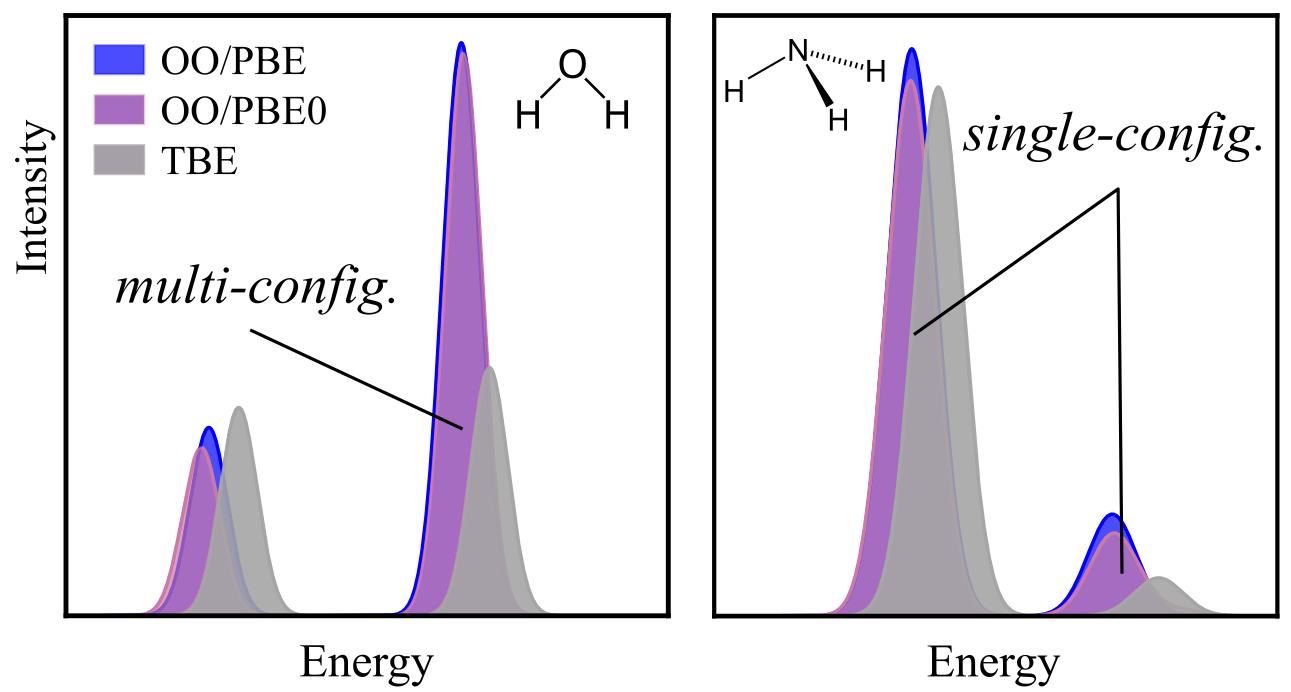}
\end{tocentry}

\bibliography{biblio.bib}
\end{document}


\tableofcontents
\clearpage

\section{Numerical Tests}

\begin{table}[h!] \caption{Differences in the excitation energy (eV) and the oscillator strength upon increasing the vacuum around the atoms from 10.5 to 15~\AA. Values are shown for the two most diffuse excited states of methanol described using different basis representations and the PBE functional.} 
\label{tab:vacum_size} 
\begin{tabular}{lccS[table-format=-1.0e-1]} 
\hline 
State & Basis & $\Delta \Delta E$ & {$\Delta f$} \\ 
\hline M$_2$ & aug-cc-pVDZ+sz & -0.001 & -8e-8 \\ 
M$_2$ & d-aug-cc-pVDZ+sz & -0.002 & -3e-7 \\ 
M$_2$ & plane waves & 0.002 & -1e-5 \\ 
M$_3$ & aug-cc-pVDZ+sz & -0.002 & -7e-6 \\ 
M$_3$ & d-aug-cc-pVDZ+sz & -0.001 & -1e-6 \\ 
M$_3$ & plane waves & 0.002 & 5e-5 \\ 
\hline 
\end{tabular} 
\end{table}

\begin{table}[h!]
\centering
\caption{LR-TDDFT calculations of the excited states of methanol at the PBE level of theory, using two different LCAO basis sets.}
\label{tab:pbe_test}
\begin{tabular}{llcccccc}
\hline
 &
   &
  \multicolumn{2}{c}{d-aug-cc-pVTZ} &
  \multicolumn{2}{c}{d-aug-cc-pVQZ} &
  \multicolumn{2}{c}{Difference} \\ \cline{3-8} 
State &
  Character &
  \multicolumn{1}{c}{$\Delta E$} &
  \multicolumn{1}{c}{$f$} &
  \multicolumn{1}{c}{$\Delta E$} &
  \multicolumn{1}{c}{$f$} &
  \multicolumn{1}{c}{$\Delta\Delta E$} &
  \multicolumn{1}{c}{$\Delta f$} \\ \hline
S$_1$ & $n \rightarrow 3$s   & 5.58 & 0.00141 & 5.59 & 0.00138 & 0.00882 & $-3.0510\times10^{-5}$ \\
S$_2$ & $n \rightarrow 3$p$_z$ & 6.17 & 0.02060 & 6.18 & 0.02047 & 0.00722 & $-1.26606\times10^{-4}$ \\
S$_3$ & $n \rightarrow 3$p$_y$ & 6.49 & 0.00647 & 6.49 & 0.00622 & 0.00581 & $-2.50264\times10^{-4}$ \\
S$_4$ & $n \rightarrow 3$p$_y$ & 6.49 & 0.00017 & 6.49 & 0.00014 & 0.00415 & $-2.5457\times10^{-5}$ \\
S$_5$ & $n^\prime \rightarrow \sigma/3$s          & 7.20 & 0.00064 & 7.21 & 0.00064 & 0.00598 & $-7.7520\times10^{-6}$ \\ \hline
\end{tabular}
\end{table}

\begin{table}[h!]
\centering
\caption{LR-TDDFT calculations of the excited states of methanol at the PBE0 level of theory, using two different LCAO basis sets.}
\label{tab:pbe0_test}
\begin{tabular}{llcccccc}
\hline
 &
   &
  \multicolumn{2}{c}{d-aug-cc-pVTZ} &
  \multicolumn{2}{c}{d-aug-cc-pVQZ} &
  \multicolumn{2}{c}{Difference} \\ \cline{3-8}
State &
  Character &
  \multicolumn{1}{c}{$\Delta E$} &
  \multicolumn{1}{c}{$f$} &
  \multicolumn{1}{c}{$\Delta E$} &
  \multicolumn{1}{c}{$f$} &
  \multicolumn{1}{c}{$\Delta\Delta E$} &
  \multicolumn{1}{c}{$\Delta f$} \\ \hline
S$_1$ & $n \rightarrow 3$s   & 6.42 & 0.00293 & 6.43 & 0.00290 & 0.00689 & $-2.3722\times10^{-5}$ \\
S$_2$ & $n \rightarrow 3$p$_z$ & 7.19 & 0.02583 & 7.20 & 0.02579 & 0.00554 & $-3.5667\times10^{-5}$ \\
S$_3$ & $n \rightarrow 3$p$_x$ & 7.60 & 0.00018 & 7.61 & 0.00015 & 0.00637 & $-2.5775\times10^{-5}$ \\
S$_4$ & $n \rightarrow 3$p$_y$ & 7.60 & 0.01234 & 7.61 & 0.01229 & 0.00670 & $-5.2303\times10^{-5}$ \\
S$_5$ & $n^\prime \rightarrow \sigma/3$s & 8.00 & 0.00066 & 8.01 & 0.00072 & 0.00448 & $5.8021\times10^{-5}$ \\ \hline
\end{tabular}
\end{table}

\clearpage

\section{Water}
\subsection{Molecular orbitals}
\begin{figure}[hbt]
    \centering
    \includegraphics[width=0.35\linewidth]{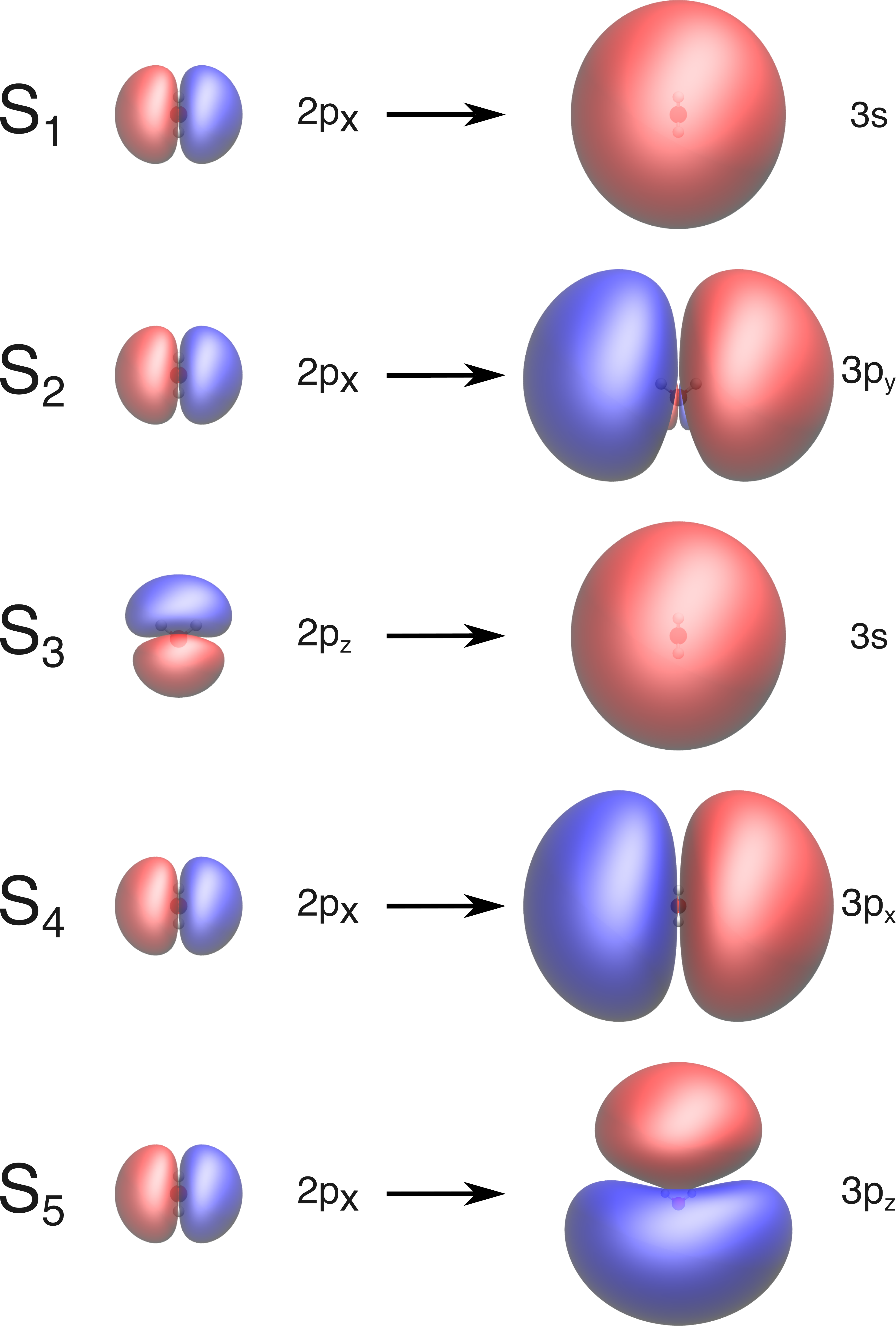}
    \caption{Molecular orbitals associated with the first five singlet Rydberg excitations of water at the Franck–Condon geometry. The orbitals shown on the left are obtained from a ground-state calculation at the PBE/\gls{pw} level of theory, while those on the right are obtained from orbital-optimized PBE/\gls{pw} calculations of the corresponding excited state.}
    \label{water_MOs}
\end{figure}

\subsection{Table of overlaps}
\begin{table}[hbt]
\centering
\caption{Overlaps between the ground state and the orbital-optimized mixed-spin excited states of water computed using different basis sets and exchange--correlation functionals.}
\label{tab:h2o_overlap}

\begin{tabular}{llccccc}
\hline
Basis/rep. & XC
& S$_1$ & S$_2$ & S$_3$ & S$_4$ & S$_5$ \\
\hline
aug-cc-pVDZ+sz   & PBE       & 0.0000 & 0.0000 & 0.0543 & 0.0225 & 0.0000 \\
d-aug-cc-pVDZ+sz & PBE       & 0.0000 & 0.0000 & 0.0516 & 0.0232 & 0.0000 \\
plane waves      & PBE       & 0.0000 & 0.0000 & 0.0226 & 0.0504 & 0.0000 \\
plane waves      & PBE0      & 0.0000 & 0.0000 & 0.0000 & 0.0166 & 0.0000 \\
plane waves      & PBE-SIC/2 & 0.0005 & 0.0000 & 0.0023 & 0.0148 & 0.0045 \\
plane waves      & PBE-SIC   & 0.0007 & 0.0004 & 0.0237 & 0.0200 & 0.0072 \\
\hline
\end{tabular}

\end{table}

\clearpage
\newpage

\section{Formaldehyde}
\subsection{Molecular orbitals}
\begin{figure}[hbt]
    \centering
    \includegraphics[width=0.35\linewidth]{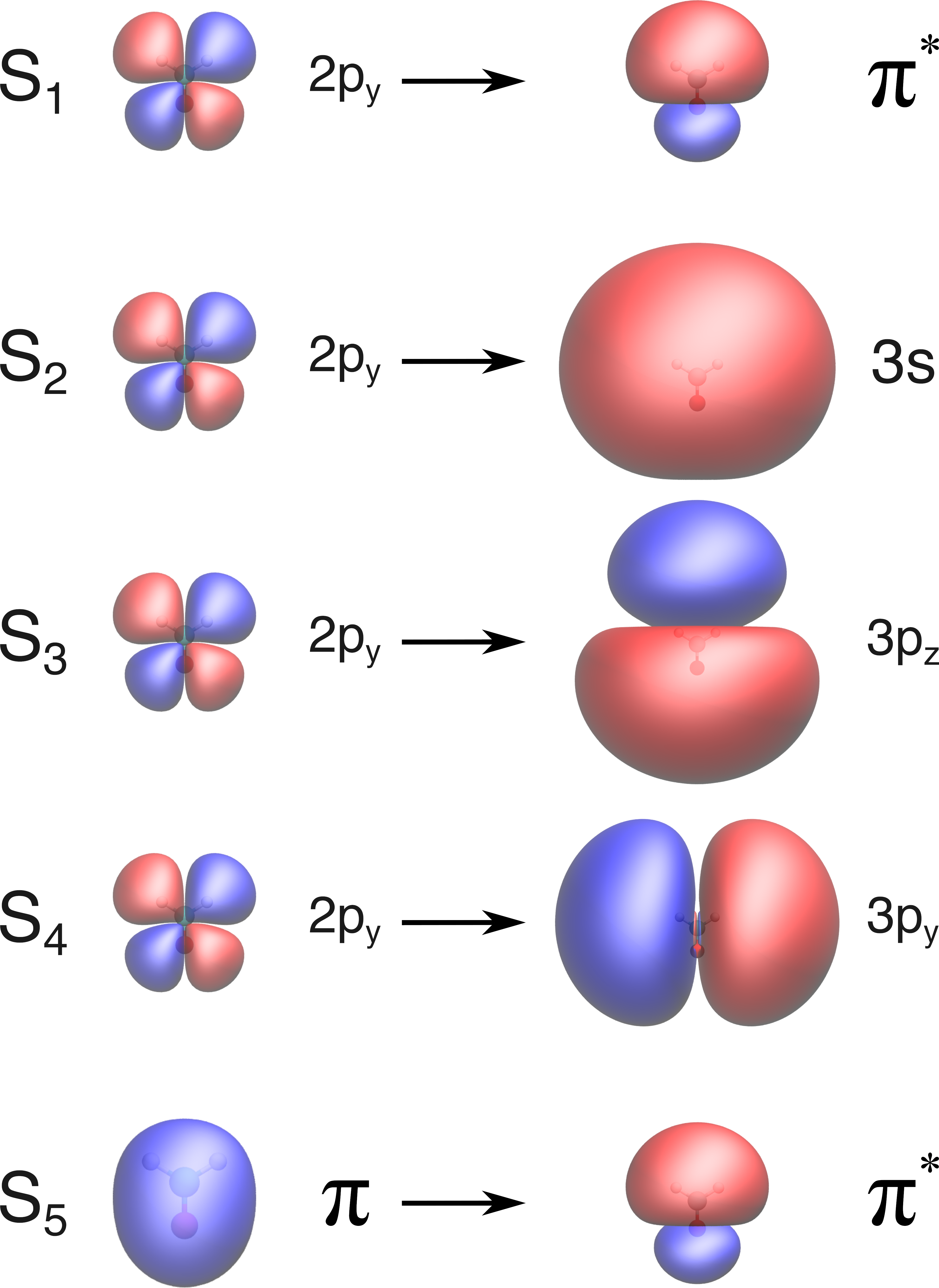}
    \caption{Molecular orbitals associated with the first five singlet excitations of formaldehyde at the Franck–Condon geometry. The orbitals shown on the left are obtained from a ground-state calculation at the PBE/\gls{pw} level of theory, while those on the right are obtained from orbital-optimized PBE/\gls{pw} calculations of the corresponding excited state.}
    \label{formaldehyde_MOs}
\end{figure}

\subsection{Table of overlaps}
\begin{table}
\centering
\caption{Overlaps between the ground state and the orbital-optimized mixed-spin excited states of formaldehyde computed using different basis sets and exchange--correlation functionals.}
\label{tab:ch2o_overlap}

\begin{tabular}{llccccc}
\hline
Basis/rep. & XC
& S$_1$ & S$_2$ & S$_3$ & S$_4$ & S$_5$ \\
\hline
aug-cc-pVDZ+sz   & PBE       & 0.0000 & 0.0000 & 0.0001 & 0.0326 & 0.0106 \\
d-aug-cc-pVDZ+sz & PBE       & 0.0000 & 0.0000 & 0.0000 & 0.0297 & 0.0104 \\
plane waves      & PBE       & 0.0000 & 0.0000 & 0.0000 & 0.0292 & 0.0084 \\
plane waves      & PBE0      & 0.0000 & 0.0000 & 0.0000 & 0.0131 & 0.0417 \\
plane waves      & PBE-SIC/2 & 0.0770 & 0.0000 & 0.0000 & 0.0210 & 0.0440 \\
plane waves      & PBE-SIC   & 0.1444 & 0.0000 & 0.0000 & 0.0184 & 0.0914 \\
\hline
\end{tabular}

\end{table}

\clearpage
\newpage

\section{Ammonia}
\subsection{Molecular orbitals}

\begin{figure}[hbt]
    \centering
    \includegraphics[width=0.35\linewidth]{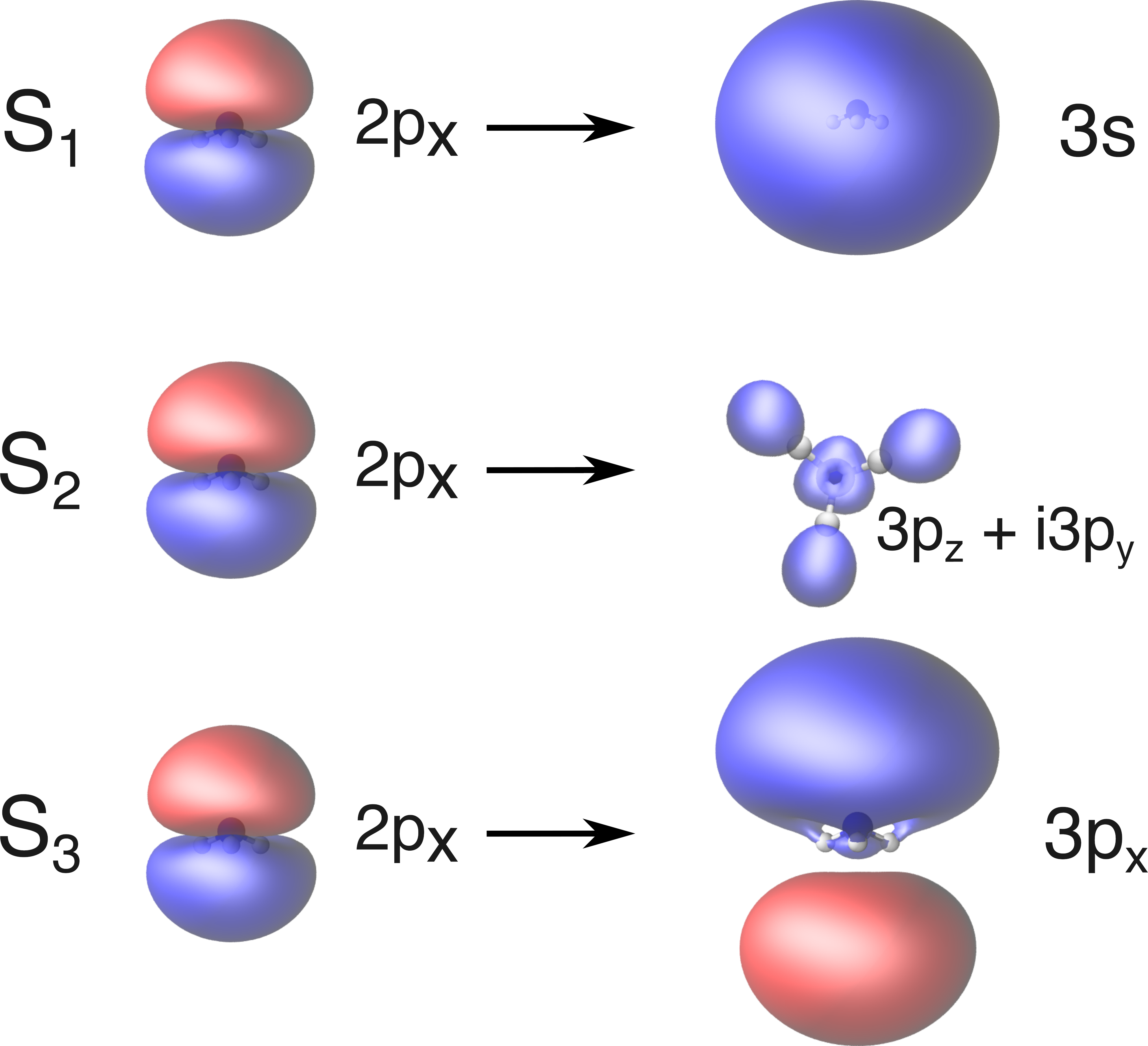}
    \caption{Molecular orbitals associated with the first three singlet excitations of ammonia at the Franck–Condon geometry. The orbitals shown on the left are obtained from a ground-state calculation at the PBE/\gls{pw} level of theory, while those on the right are obtained from orbital-optimized PBE/\gls{pw} calculations of the corresponding excited state. For S$_2$ with symmetry $E$, the orbital density of the linear combination $\mathrm{3p}_z + i3\mathrm{p}_y$ is shown instead.}
    \label{ammonia_MOs}
\end{figure}

\subsection{Table of overlaps}
\begin{table}[hbt]
\centering
\caption{Overlaps between the ground state and the orbital-optimized mixed-spin excited states of ammonia computed using different basis sets and exchange--correlation functionals.}
\label{tab:nh3_overlap}

\begin{tabular}{llccc}
\hline
Basis/rep. & XC
& S$_1$ & S$_2$ & S$_3$ \\
\hline
aug-cc-pVDZ+sz   & PBE       & 0.0153  & 0.0000 & 0.0522 \\
d-aug-cc-pVDZ+sz & PBE       & 0.0152  & 0.0000 & 0.0513 \\
plane waves      & PBE       & 0.0146 & 0.0000 & 0.0501 \\
plane waves      & PBE0      & 0.0011 & 0.0000 & 0.0173 \\
plane waves      & PBE-SIC/2 & 0.2657  & 0.0248 & 0.0020 \\
plane waves      & PBE-SIC   & 0.2438  & 0.0130 & 0.0036 \\
\hline
\end{tabular}

\end{table}

\clearpage
\newpage

\section{Methanol}
\subsection{Molecular orbitals}

\begin{figure}[hbt]
    \centering
    \includegraphics[width=0.35\linewidth]{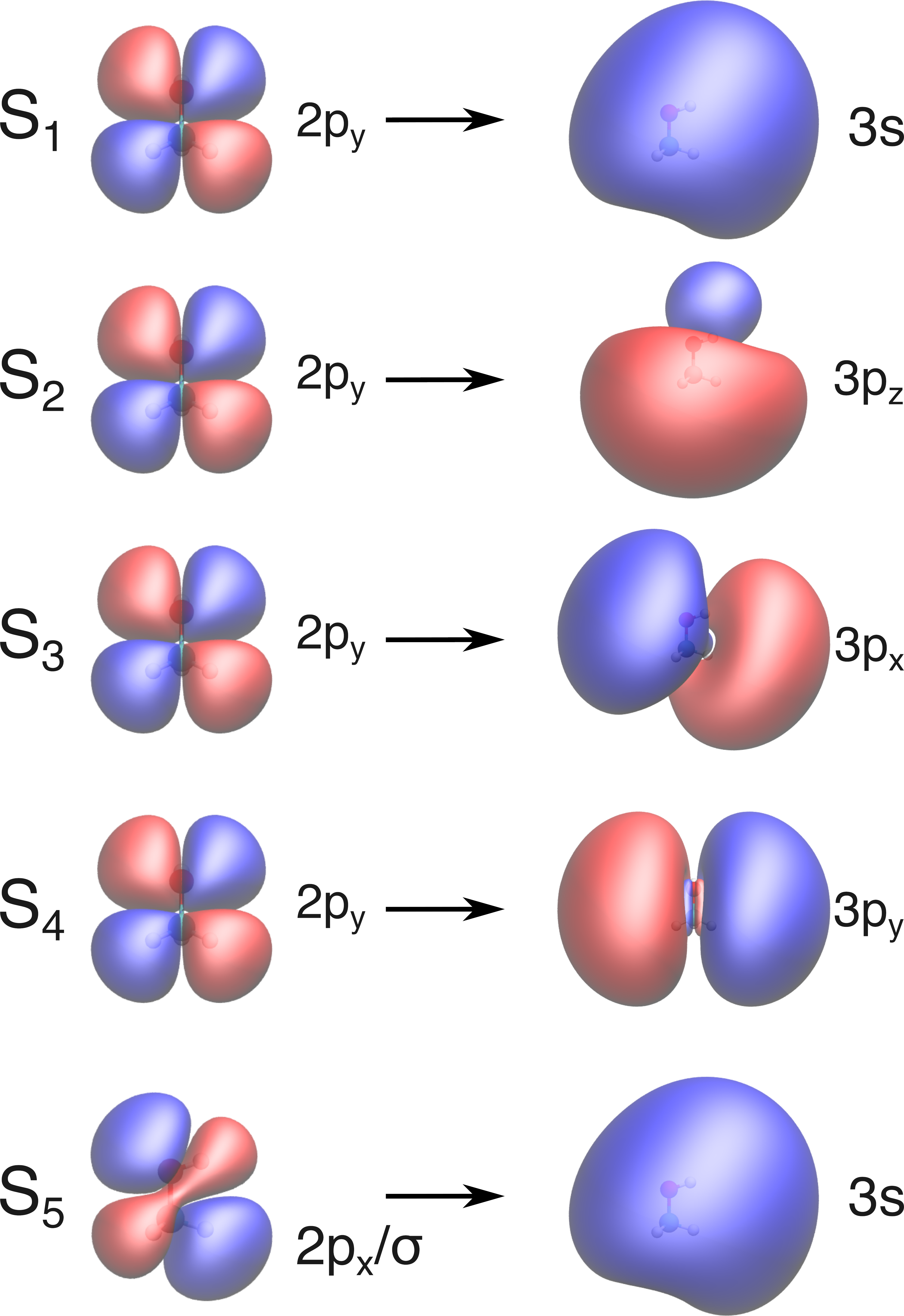}
    \caption{Molecular orbitals associated with the first five singlet excitations of methanol at the Franck–Condon geometry. The orbitals shown on the left are obtained from a ground-state calculation at the PBE/\gls{pw} level of theory, while those on the right are obtained from orbital-optimized PBE/\gls{pw} calculations of the corresponding excited state.}
    \label{ammonia_MOs}
\end{figure}

\subsection{Table of overlaps}
\begin{table}[hbt]
\centering
\caption{Overlaps between the ground state and the orbital-optimized mixed-spin excited states of methanol computed using different basis sets and exchange--correlation functionals.}
\label{tab:methanol_overlap}

\begin{tabular}{llccccc}
\hline
Basis/rep. & XC
& S$_1$ & S$_2$ & S$_3$ & S$_4$ & S$_5$ \\
\hline
aug-cc-pVDZ+sz   & PBE       & 0.0000 & 0.0000 & 0.0000 & 0.0152 & 0.0025 \\
d-aug-cc-pVDZ+sz & PBE       & 0.0000 & 0.0000 & 0.0000 & 0.0029 & 0.0025 \\
plane waves      & PBE       & 0.0000 & 0.0000 & 0.0000 & 0.0025 & 0.0026 \\
plane waves      & PBE0      & 0.0000 & 0.0000 & 0.0000 & 0.0038 & 0.0061 \\
plane waves      & PBE-SIC/2 & 0.0027 & 0.0029 & 0.0015 & 0.0004 & 0.0104 \\
plane waves      & PBE-SIC   & 0.0115 & 0.0043 & 0.0082 & 0.0063 & 0.0121 \\
\hline
\end{tabular}

\end{table}

\clearpage
\newpage

\section{Ethylene}
\subsection{Molecular orbitals}

\begin{figure}[hbt]
    \centering
    \includegraphics[width=0.35\linewidth]{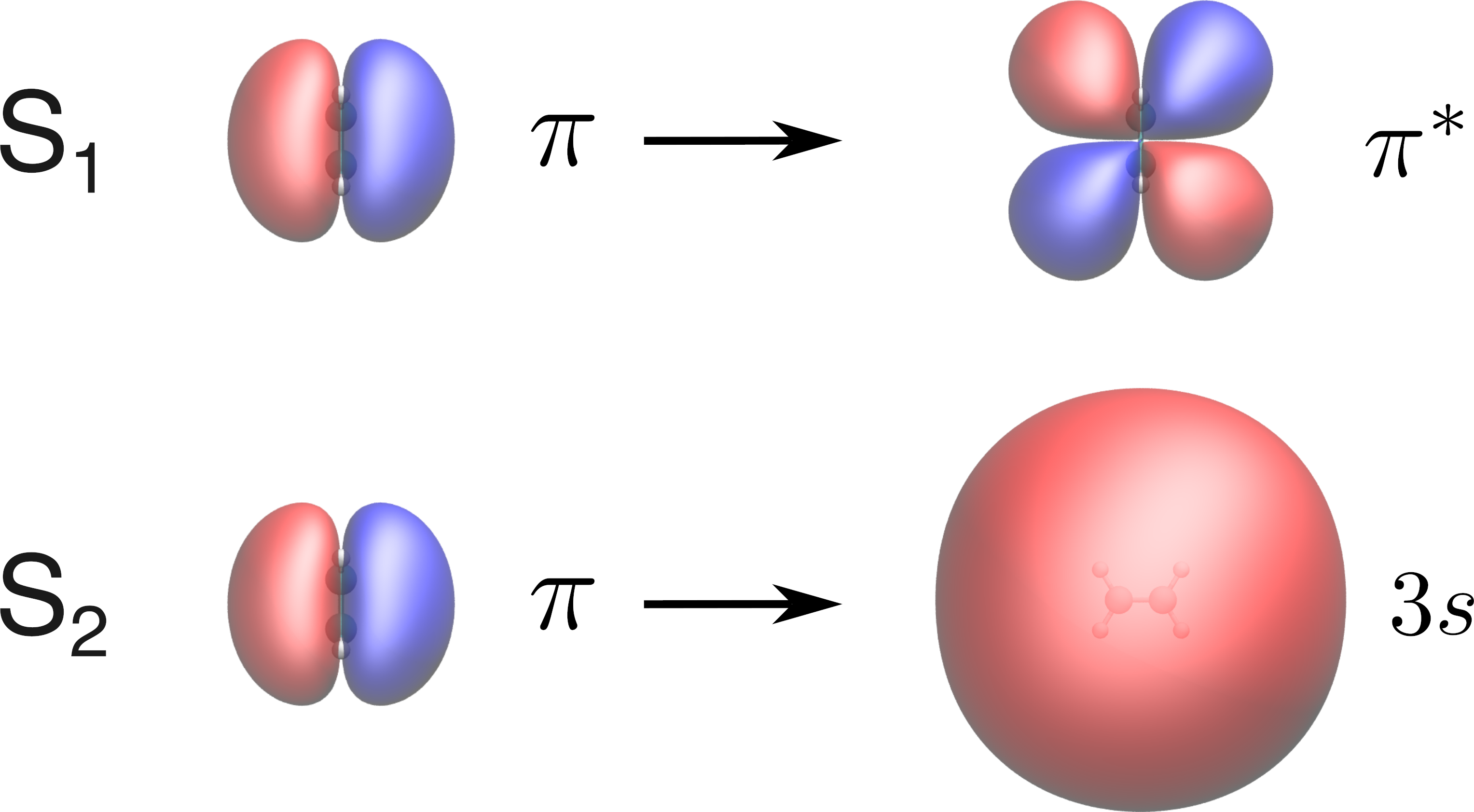}
    \caption{Molecular orbitals associated with two singlet excitations of ethylene at the Franck–Condon geometry. The orbitals shown on the left are obtained from a ground-state calculation at the PBE/\gls{pw} level of theory, while those on the right are obtained from orbital-optimized PBE/\gls{pw} calculations of the corresponding excited state.}
    \label{ethylene_MOs}
\end{figure}

\subsection{Table of overlaps}
\begin{table}[hbt]
\centering
\caption{Overlaps between the ground state and the orbital-optimized mixed-spin excited states of ethylene computed using different basis sets and exchange--correlation functionals.}
\label{tab:c2h4_pure_singlets_overlap}

\begin{tabular}{llcc}
\hline
Basis/rep. & XC
& S$_1$ & S$_2$ \\
\hline
aug-cc-pVDZ+sz   & PBE       & 0.0382 & 0.0000 \\
d-aug-cc-pVDZ+sz & PBE       & 0.0380 & 0.0000 \\
plane waves      & PBE       & 0.0374 & 0.0000 \\
plane waves      & PBE0      & 0.0166 & 0.0000 \\
plane waves      & PBE-SIC/2 & 0.0084 & 0.0000 \\
plane waves      & PBE-SIC   & 0.0045 & 0.0000 \\
\hline
\end{tabular}

\end{table}

\clearpage
\newpage

\section{Additional plots}

\begin{figure}[hbt]
    \centering
    \includegraphics[width=\linewidth]{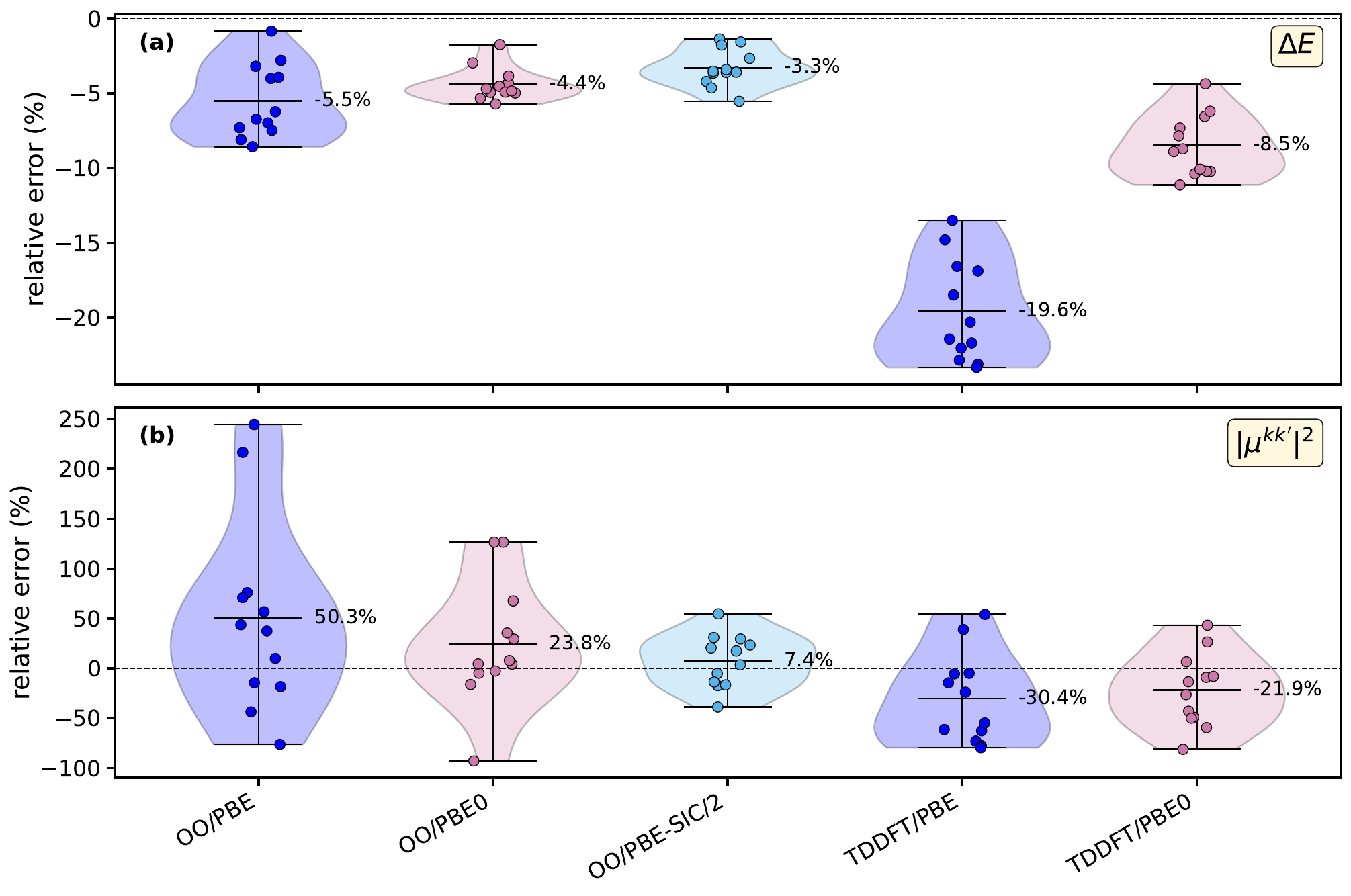}
    \caption{Signed relative errors on (a) the excitation energy and (b) the squared magnitude of the transition dipole moment for orbital-optimized and linear-response time-dependent density functional theory calculations using various exchange-correlation functionals. The mean value is indicated by a horizontal bar. Only single-configurational  states with $f_{\mathrm{ref}} > 0.001$ are included.}
    \label{fig:error_E_TDM_SC}
\end{figure}

\begin{figure}[hbt]
    \centering
    \includegraphics[width=\linewidth]{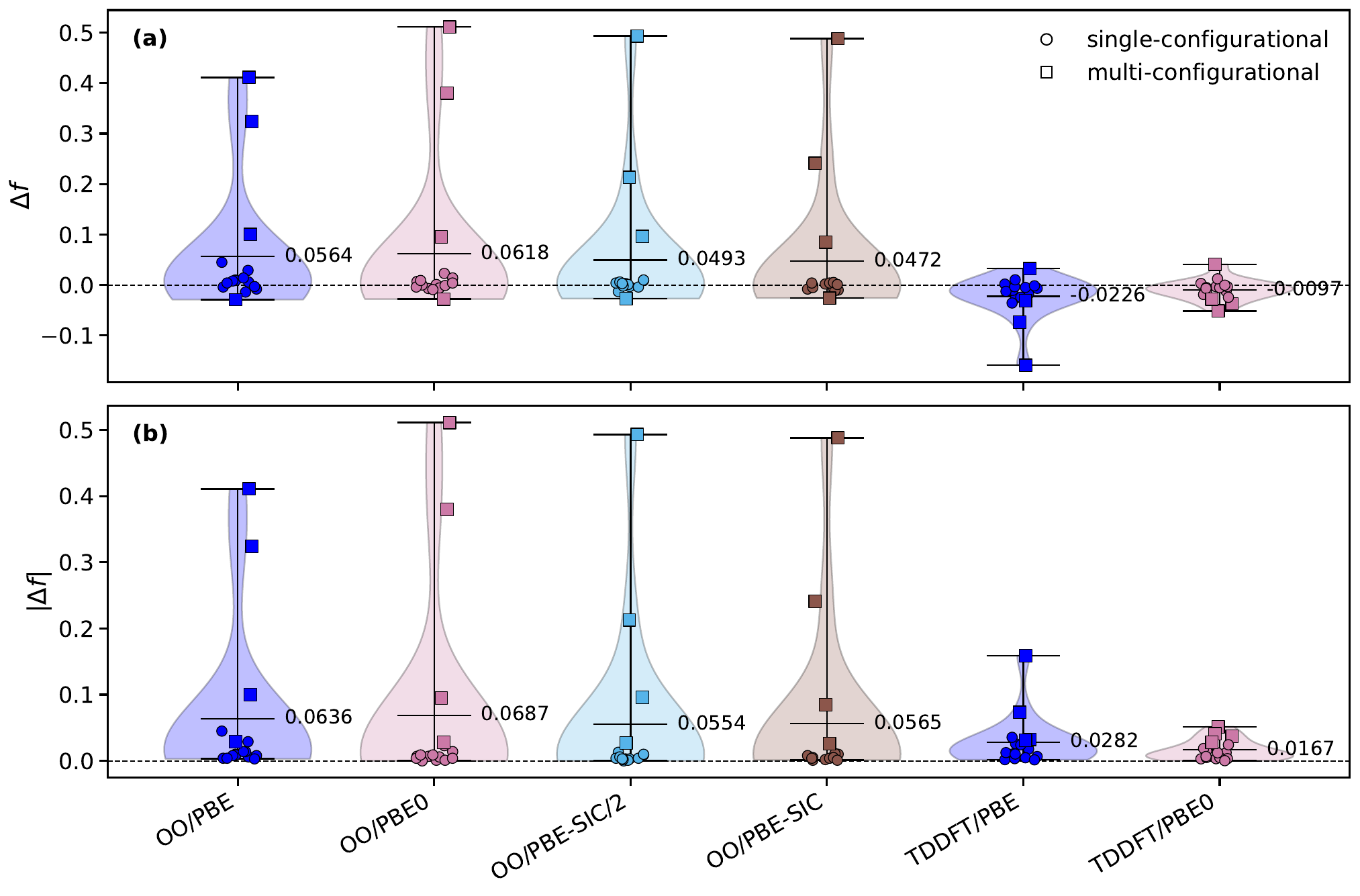}
    \caption{
    Signed errors (a) and absolute errors (b) in oscillator strength for orbital-optimized (OO) and linear-response time-dependent density functional theory (LR-TDDFT) calculations using various exchange-correlation functionals. The mean value is indicated by a horizontal bar. Only states with $f_{\mathrm{ref}} > 0.001$ are included. Excited states with multi-configurational character in higher-level calculations from the literature are marked with squares, while those with single-configurational character are marked with circles. The OO calculations exhibit much larger errors for multi-configurational states than LR-TDDFT, independently of the functional.}
    \label{fig:abs_error}
\end{figure}

\begin{figure}[hbt]
    \centering
    \includegraphics[width=\linewidth]{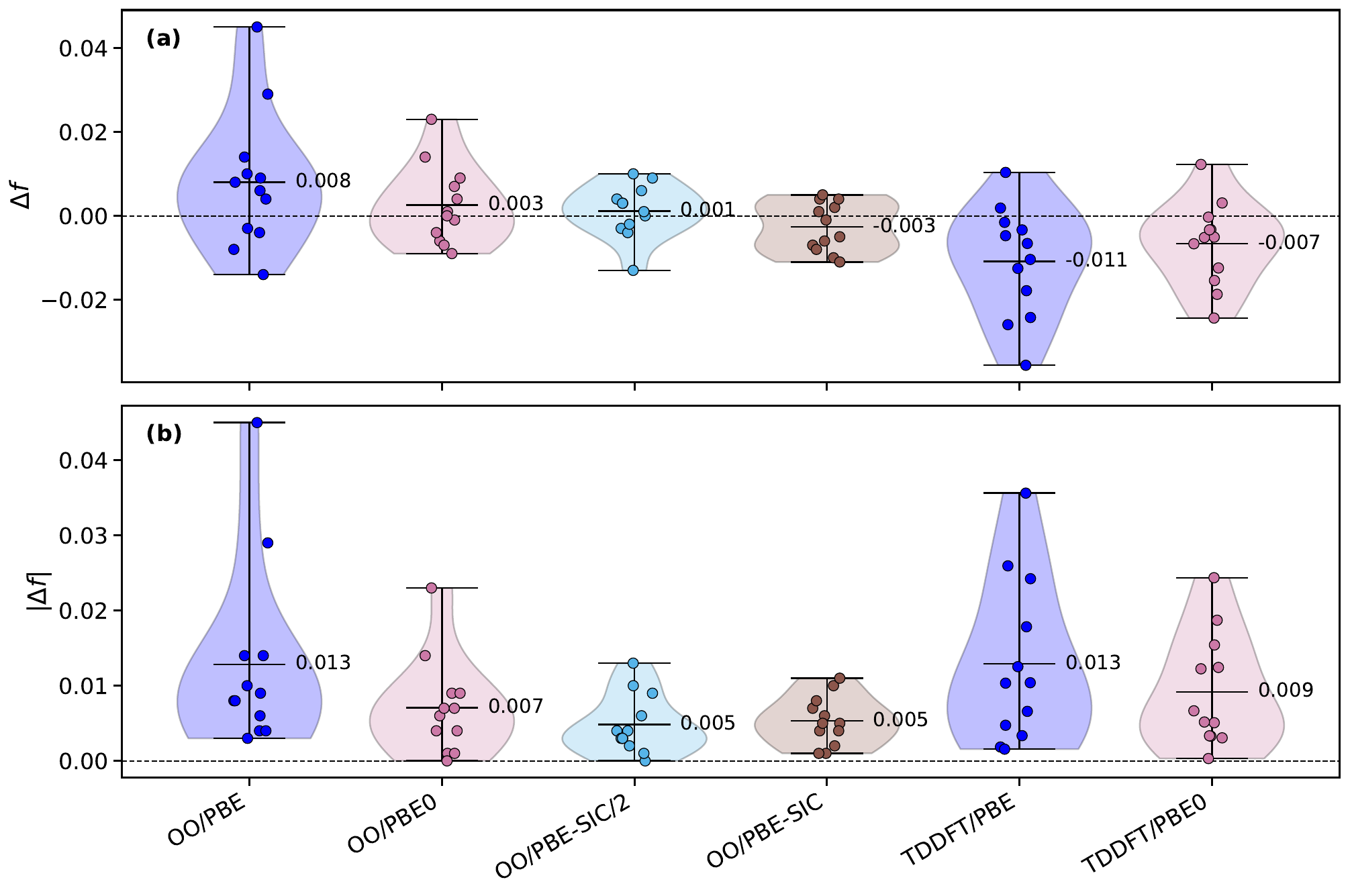}
    \caption{Signed errors (a) and absolute errors (b) in oscillator strength for orbital-optimized and linear-response time-dependent density functional theory for various exchange-correlation functional. The mean value is indicated as a horizontal bar. Only states with $f_{\mathrm{ref}} > 0.001$ and a single-configurational character in higher-level calculations are included.}
    \label{fig:relative_error_no_multi}
\end{figure}

\begin{figure}[hbt]
    \centering
    \includegraphics[width=\linewidth]{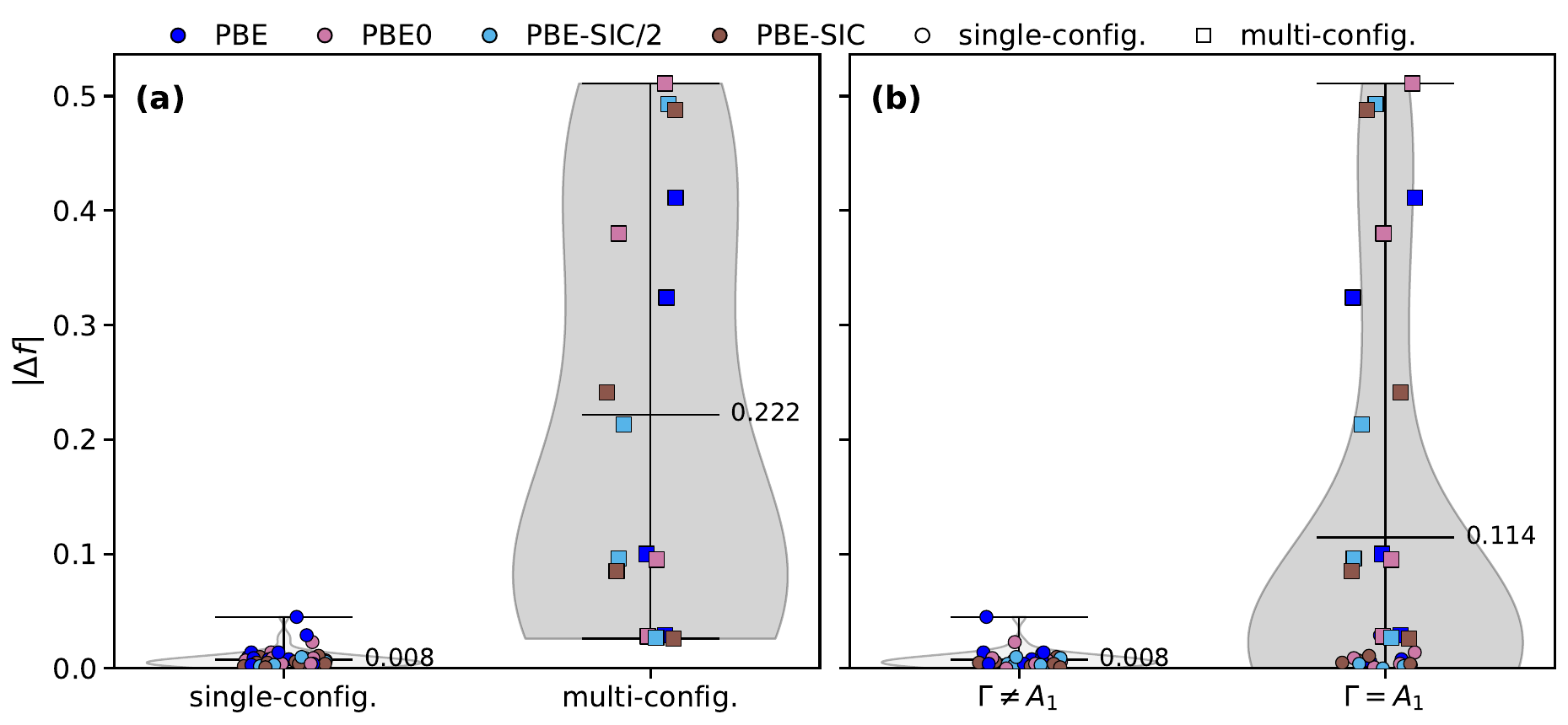}
    \caption{Absolute errors in oscillator strength from orbital-optimized calculations, grouped according to (a) the single- or multi-configurational character of the excited states and (b) their irreducible representation relative to that of the ground state. States belonging to the totally symmetric representation ($A_1$) have nonzero overlap with the ground state. The mean value is indicated by a horizontal bar. Only states with $f_{\mathrm{ref}} > 0.001$ are included. The largest errors are associated primarily with multi-configurational states rather than with nonzero overlap between the ground and excited state.}
    \label{fig:overlap_error}
\end{figure}

\begin{figure}[hbt]
    \centering
    \includegraphics[width=\linewidth]{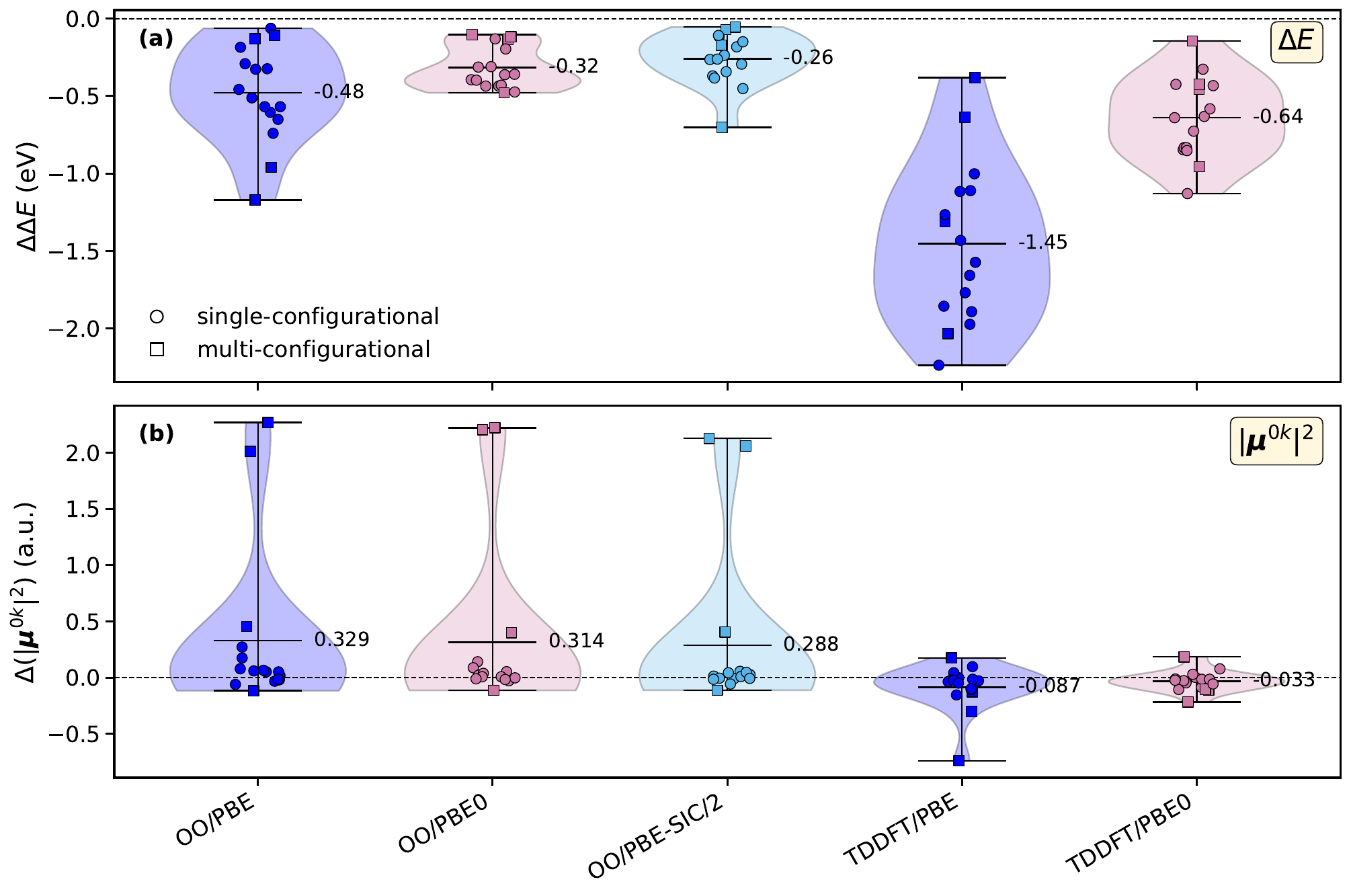}
    \caption{Signed errors on (a) the excitation energy and (b) the squared magnitude of the transition dipole moment for orbital-optimized (OO) and linear-response time-dependent density functional theory (LR-TDDFT) calculations using various exchange-correlation functionals. The mean value is indicated by a horizontal bar. Only states with $f_{\mathrm{ref}} > 0.001$ are included. Excited states with multi-configurational character are marked with squares, those with single-configurational character with circles. The OO calculations are affected by smaller errors on the excitation energy, but larger errors on the transition dipole moment, especially for the multi-configurational states.}
    \label{fig:error_energy_vs_TDM}
\end{figure}
